\documentclass[journal]{IEEEtran}
\usepackage{amsmath,amssymb,amsfonts}
\usepackage{mathtools}
\usepackage{algorithmic}
\usepackage{array}
\usepackage[caption=false,font=normalsize,labelfont=rm,textfont=sf]{subfig}
\usepackage{textcomp}
\usepackage{stfloats}
\usepackage{url}
\usepackage{verbatim}
\usepackage{graphicx}
\usepackage{cite}
\def\BibTeX{{\rm B\kern-.05em{\sc i\kern-.025em b}\kern-.08em
    T\kern-.1667em\lower.7ex\hbox{E}\kern-.125emX}}
\usepackage{balance}
\usepackage{xcolor}
\usepackage{booktabs}
\usepackage{hyperref}

\begin{document}
\title{Experimental Performance of Bidirectional Phase Coherent Transmission and Sensing for mmWave Cell-free Massive MIMO Systems with Reciprocity Calibration}
\author{Qingji Jiang, Jing jin, Qixing Wang, Yuanyuan Tang, Yang Cao, Bin Kuang, Jing Dong, Siying Lv, Dongming Wang, Yongming Huang, Jiangzhou Wang, Xiaohu You

\thanks{Qingji Jiang, Yuanyuan Tang are with the National Mobile Communications Research Laboratory, Southeast University, Nanjing 210096, China (email, jiangqingji@seu.edu.cn;)}
\thanks{Bin Kuang is with the Pervasive Communication Research Center, Purple Mountain Laboratories, Nanjing 211111, China.}
\thanks{Jing jin, Qixing Wang, Jing Dong, Siying Lv are with the China Mobile Research Institute, Beijing 100053, China.}
\thanks{Yang Cao, Dongming Wang, Yongming Huang, Jiangzhou Wang, Xiaohu You are with the National Mobile Communications Research Laboratory, Southeast University, Nanjing 210096, China, and also with the Pervasive Communication Research Center, Purple Mountain Laboratories, Nanjing 211111, China. (e-mail: wangdm@seu.edu.cn). (Corresponding author,  Dongming Wang.)}}
\markboth{Journal of \LaTeX\ Class Files,~Vol.~18, No.~9, September~2020}%
{Experimental Performance of Bidirectional Phase Coherent Transmission and Sensing for mmWave Cell-free Massive MIMO Systems with Reciprocity Calibration}

\maketitle

\begin{abstract}
Phase synchronization among distributed transmission reception points (TRPs) is a prerequisite for enabling coherent joint transmission and high-precision sensing in millimeter wave (mmWave) cell-free massive multiple-input and multiple-output (MIMO) systems. This paper proposes a bidirectional calibration scheme and a calibration coefficient estimation method for phase synchronization, and presents a calibration coefficient phase tracking method using unilateral uplink/downlink channel state information (CSI). Furthermore, this paper introduces the use of reciprocity calibration to eliminate non-ideal factors in sensing and leverages sensing results to achieve calibration coefficient phase tracking in dynamic scenarios, thus enabling bidirectional empowerment of both communication and sensing. Simulation results demonstrate that the proposed method can effectively implement reciprocal calibration with lower overhead, enabling coherent collaborative transmission, and resolving non-ideal factors to acquire lower sensing error in sensing applications. Experimental results show that, in the mmWave band, over-the-air (OTA) bidirectional calibration enables coherent collaborative transmission for both collaborative TRPs and collaborative user equipments (UEs), achieving beamforming gain and long-time coherent sensing capabilities.
\end{abstract}

\begin{IEEEkeywords}
 millimeter wave cell-free massive MIMO, reciprocity calibration, coherent joint transmission and sensing, collaborative UE, prototype system.
\end{IEEEkeywords}

\section{Introduction}
\IEEEPARstart{C}{ell-free} massive multiple-input multiple-output (CF-mMIMO) technology, or cooperative techniques with large-scale distributed nodes, not only enhance the system's spectral efficiency, peak rate, and reliability, but also improve positioning and sensing accuracy\cite{cao2025}. As a result, CF-mMIMO is regarded as a key technology for the sixth generation mobile communication (6G) by both the academic and industrial communities.

CF-mMIMO collaborative technology places extremely high demands on phase synchronization among distributed nodes. Timing offsets, frequency offsets, and phase jitter between different nodes can all affect synchronization across the distributed nodes\cite{Xin2025}. In the sensing literature, these deviation terms are often referred to as non-ideal factors. Therefore, to achieve cooperative communication, it is necessary to acquire the downlink channel for precoding, compensate for the phase deviations between different nodes, and thereby realize cooperative communication. For cooperative sensing, the timing and frequency offsets existing among multiple nodes must be calculated and eliminated.

However, as the number of distributed nodes increases, the cost of obtaining downlink channel through feedback becomes unacceptably high\cite{Xu2024}. Fortunately, thanks to channel reciprocity, the physical wireless channel remains identical for both uplink and downlink within the channel coherence time. Therefore, the uplink channel can be directly utilized to compute the downlink channel, thereby avoiding the burdensome overhead of pilot feedback. In time division duplex (TDD) systems, downlink (DL) precoding can be performed based on uplink (UL) channel state information (CSI) by leveraging channel reciprocity, thereby enabling  DL coherent transmission \cite{Smith2004}. This reciprocity-based approach underpins various technologies including massive multiple-input and multiple-output (MIMO) \cite{3}, coordinated multi-point (CoMP) \cite{4}, multiple transmission and reception point (Multi-TRP), coherent joint transmission (CJT) and CF-mMIMO \cite{5}.  Beamforming \cite{6} or precoding \cite{7,8} can be employed to achieve a trade-off between diversity and multiplexing, thereby enhancing system performance. These precoding strategies can be broadly categorized as codebook-based or non-codebook-based. In codebook-based approaches, the precoding matrix is selected from a predefined set based on the quantized CSI, offering robustness against channel variations \cite{9}. In contrast, non-codebook-based techniques compute the precoding matrix directly from real-time CSI, achieving superior system performance \cite{10}. The coverage of mobile communication systems is constrained by the transmission power of the user equipment (UE). Similarly, informed CSI at the UE enables effective beamforming and precoding. Therefore, UL CJT can significantly improve the performance of UE at the cell boundary. When multiple UEs can share the transmitted information, they form collaborative UEs, which can significantly enhance the UL transmission capacity. Collaborative UE technology is also considered a potential technology for 6G\cite{Kumar2025}.

However, in practice, reciprocity seldom holds due to a multitude of hardware-induced impairments. In realistic multi-node cooperative transmission systems, two dominant factors degrade reciprocity: (i) mismatched radio-frequency (RF) front-ends at the transceivers, which render the composite channel non-reciprocal even if the propagation medium itself is reciprocal, and (ii) timing and frequency offsets among distributed nodes arising from independent local oscillators (LO). Consequently, the phase-synchronization requirement for coherent cooperative communications is fundamentally translated into the problem of accurately calibrating the reciprocity of downlink and uplink channels. Prior studies \cite{12} pointed out that RF mismatches at the UE side have limited impact on DL precoding performance. Thus, compensating for the base station (BS)-side antenna mismatches alone can sufficiently restore reciprocity. The concept of self-calibration was initially proposed in the Argos method \cite{13}, where the calibration coefficients are estimated directly from inter-antenna channel measurements. Although straightforward, the Argos method is highly sensitive to channel quality. To improve robustness, several least-squares (LS) based estimators suitable for distributed systems were proposed in \cite{14}, leveraging all bidirectional inter-BS channels to achieve better performance under low signal-to-noise ratio (SNR) by exploiting spatial diversity. However, as the antenna array expands, Argos and LS methods require substantial pilot resources. To mitigate this, \cite{15} proposed using the already calibrated BS antenna array to assist in calibrating uncalibrated antennas, enabling larger arrays to be calibrated with the same pilot budget. Additionally, \cite{16} proposed tracking variations in the calibration coefficient matrix across time intervals, and exploits matrix sparsity to further reduce the pilot resource expenditure. Besides, \cite{17} proposed a spanning-tree based approach that optimizes calibration order and significantly reduces computational cost.  Self-calibration at the BS is transparent to the UE, but incurs service disruption due to the bidirectional transmission of calibration signals among BS antennas. Furthermore, under frequency offsets among remote radio units (RRUs) and time-varying channels, self-calibration must be conducted frequently, significantly increasing calibration overhead. Another method of calibration is UE-assisted calibration. In [18], the BS transmits a DL reference signal, the UE estimates the DL CSI and transmits a feedback signal, enabling the BS to calibrate RF mismatches using both UL and DL CSI. For example, \cite{19} formulated a total-least-square (TLS) problem by minimizing the error between the estimated and true channel responses, thereby deriving the reciprocity calibration coefficients. In contrast, \cite{20} constructed a TLS problem based on the principle that the UL channel and the calibrated DL channel should exhibit reciprocity, and derives the calibration coefficients accordingly. However, when the number of BS antennas is large, CSI feedback imposes a significant overhead even with space-frequency domain compression DFT codebook in the 3rd Generation Partnership Project (3GPP) protocol. To alleviate this, \cite{Qualcomm2024} proposed  estimating reciprocity calibration coefficients via precoded pilot transmission, thus avoiding CSI feedback.

As frequency bands evolve into the higher millimeter-wave spectrum, the emergence of hybrid digital-analog (HBF) arrays has increased the overhead and complexity of calibration. A typical HBF array comprises a digital precoder, digital RF chains, an analog precoder, and RF chains.
The hybrid analog-digital beamforming (HAD-BF) is a commonly used architecture in the industry for millimeter wave (mmWave)  CF-mMIMO systems. The work in \cite{23} treats the HAD-BF system as a virtual full-digital system and applies a self-calibration algorithm at the base station for antenna calibration. However, this method imposes substantial overhead for estimating equivalent uplink and downlink channels in the HBF system. Specifically, the analog precoder must exhaustively transmit a number of analog beams no fewer than the number of antennas, while the digital precoder needs to transmit beams at least equal to the number of RF chains. To this end, \cite{24} proposed a hierarchical-absolute calibration (HAC) method, decomposing the reciprocity calibration problem into digital and analog sub-chains. Similarly, \cite{25} reduced the number of pilots by replacing the digital chain calibration matrix with precomputed values during analog chain calibration.

For single-carrier systems, the aforementioned calibration algorithms assume that the mismatch coefficients remain constant over periods of hours or even days. However, in orthogonal frequency division multiplexing (OFDM) systems, when timing and frequency offsets between different LOs are considered, these mismatch coefficients may vary at a sub-second level. Consequently, when applying calibration algorithms to OFDM systems, the calibration must be performed on a per-subcarrier basis to ease the distinct phase differences. On the other hand, even if the over-the-air (OTA) channel remains static, calibration still needs to be frequently performed to restore channel reciprocity. Hence, efficient reciprocity calibration are crucial for CF-mMIMO systems\cite{Cao2023}, especially when TRPs are geographically distributed with independent LOs.

From the perspective of cooperative sensing, whether using multi-TRP collaborative sensing or UE-assisted sensing, there are higher requirements for time-frequency and phase synchronization between the nodes. Cooperative sensing can be achieved in CF-mMIMO systems\cite{Tang2025}. The asynchronous effects can be suppressed or eliminated by performing division or conjugate multiplication operations on the CSI of multiple receive antennas sharing a common clock source. The cross-array signal ratio (CASR) method \cite{27} divided the CSI of each receiving antenna by that of the reference antenna, and applied a Moore-Penrose decomposition to cancel out phase noise caused by clock asynchrony among antennas. The cross-array cross-correlation (CACC) method \cite{26} mitigated random phase shifts, timing offset (TO), and carrier frequency offset (CFO) by computing the conjugate multiplication of the CSI between each receiving antenna and a reference antenna. However, both methods rely on shared LO and clock sources, limiting their applicability in distributed systems. To address this, \cite{28} proposed a reference-path-based compensation algorithm, but its dependence on environment-specific refence paths restricts generalization. To mitigate this environmental dependency, round-trip measurements can be utilized to eliminate sensing non-ideal factors\cite{Dwivedi2015}. From the perspective of multi-node timing and frequency offset estimation, communication reciprocity calibration essentially represents one implementation of round-trip measurement. Consequently, we can leverage the reciprocity calibration coefficients across multiple OFDM subcarriers to calculate and eliminate timing and frequency offsets between distributed nodes. 

Motivated by these challenges, this paper proposes a bidirectional calibration scheme that enables time-frequency synchronization between the BS and the UE, applied to both coherent transmission and sensing for mmWave CF-mMIMO systems. The bidirectional calibration scheme involves mutual transmission and reception of reference signals between the BS antenna ports and the UE antenna ports. Furthermore, a reciprocity-based calibration coefficient tracking scheme based on UL and DL reference signals is proposed. The tracking scheme only requires one-way reception of reference signals to track the LO phase variations. This scheme is also suitable for mmWave HAD-BF systems and can be widely applied in large-scale MIMO array systems. The main contributions of this work can be summarized as follows:

\begin{itemize}
\item{{\bf Bidirectional calibration scheme.} We propose a bidirectional calibration scheme that enables the simultaneous acquisition of calibration coefficients for both BS and UE, thereby realizing coherent cooperative transmission on both BS and UE sides. Additionally, a calibration coefficient phase tracking scheme is proposed, which can utilize unilateral UL/DL CSI to track phase variations of the calibration coefficients, helping to reduce the calibration overhead.}
\item{{\bf Calibration-assisted sensing.}  We introduce the method of 
calibration-assisted sensing, where reciprocity calibration is leveraged to eliminate 
timing and frequency offsets across distributed nodes, effectively mitigating key non-ideal 
factors and enhancing sensing accuracy. Experiments validate that calibration-assisted sensing can help realize long-time coherent sensing. }
\item{{\bf Sensing-assisted calibration.} We propose a phase tracking scheme for reciprocity calibration coefficients to address calibration in dynamic 
scenarios. Environmental sensing is leveraged to reconstruct the OTA channel, thereby suppressing dynamic phase disturbances and enhancing the robustness of calibration coefficient phase tracking.Simulation results show that it can reduce the calibration overhead to 1/10 while remaining coherent transmission.}
\end{itemize}

The paper is organized as follows: Section \ref{section2} introduces the reciprocity modeling of mmWave CF-mMIMO systems. Section \ref{section3} presents the bidirectional calibration scheme in a quasi-static scenario. Section \ref{section4} discusses calibration-assisted sensing and sensing-assisted calibration schemes in dynamic scenarios. Section \ref{section5} provides experimental verification results. Section \ref{section6} concludes the paper.

Matrices and vectors are denoted by bold uppercase and lowercase letters, respectively; 
$\mathbf{G}(k,l)$ denotes the element of the matrix; $\mathbf{G}_{m,n}$ denotes the matrix; 
$(\cdot)^{H}$, $(\cdot)^{T}$, and $(\cdot)^{*}$ represent Hermitian transpose, transpose, and conjugate, respectively; $\text{diag}(x)$ is a diagonal matrix with $x$ on its diagona; and $\otimes$ denotes the Kronecker product of two matrices.

\section{System Model}
\label{section2}
\subsection{System Modelling}
\begin{figure}[htbp]
\centering
\includegraphics[width=0.65\linewidth]{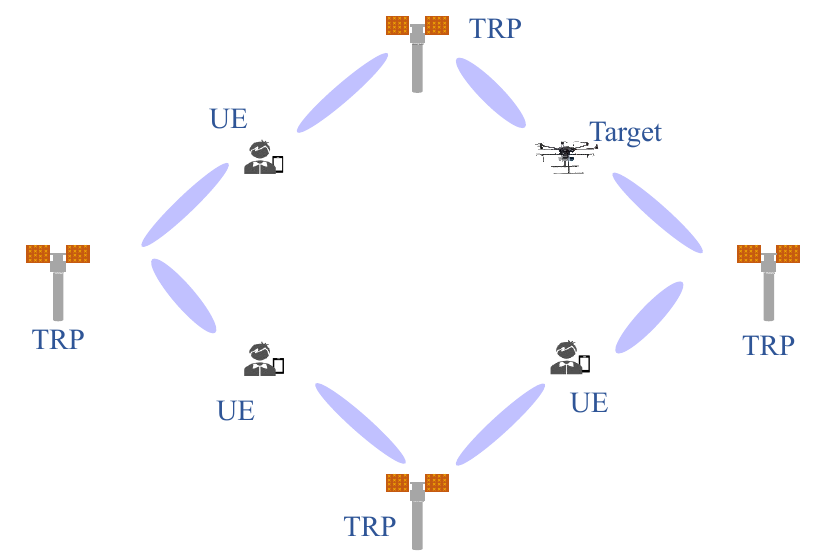}
\caption{Diagram for mmWave cell-free massive MIMO system.}
\label{fig_m2m}
\end{figure}

We consider a mmWave CF-mMIMO system as shown in Fig.\ref{fig_m2m}. The system employs ${M}$  mmWave HAD-BF TRPs at the BS side to serve ${N}$ UEs on the same time-frequency resource. For the DL, multiple TRPs send multiple data streams to multiple UEs through collaborative digital-analog hybrid precoding. For the UL, multiple terminals send the same data 
stream through coherent collaborative transmission, which can enhance the coverage and realize the long-distance transmission of mmWave. On the other hand, in order to realize high-precision sensing, it is also necessary to consider the collaboration of multiple TRPs as well as UEs.

\subsection{Channel Reciprocity Modeling in mmWave Point-to-Point Systems}

\begin{figure}[t]
\centering
\includegraphics[width=0.9\linewidth]{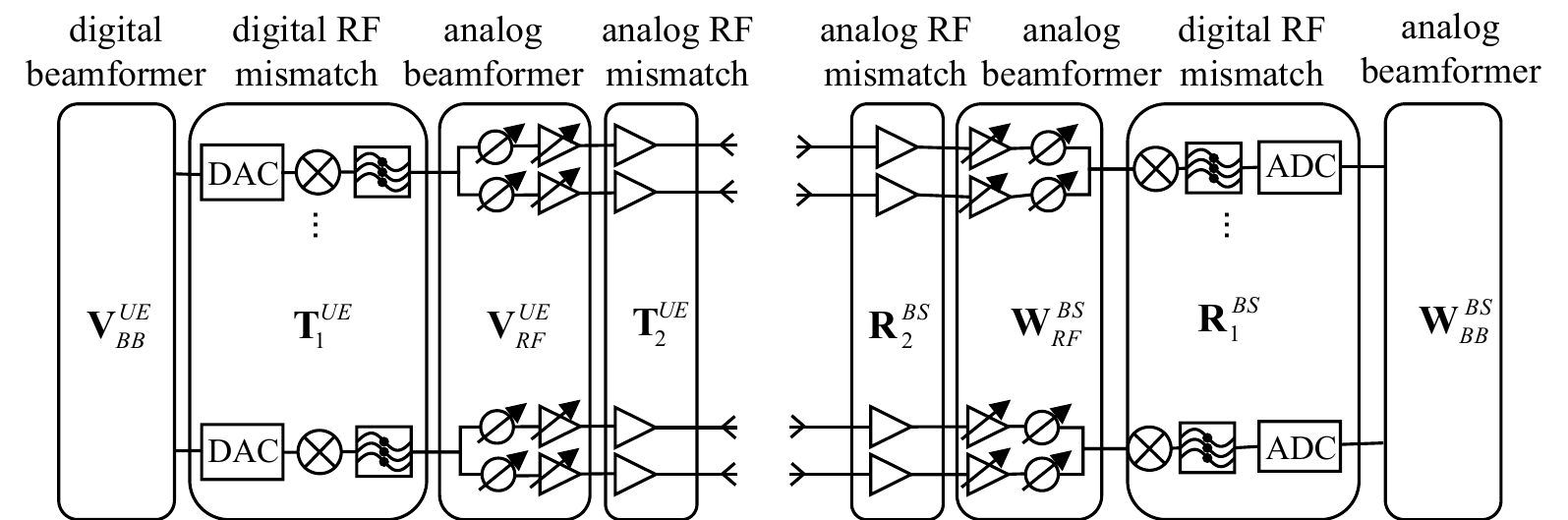}
\caption{HAD-BF system where a UE is transmtting to a BS. It can also be seen as an analog phased array when $N_{RF}^{BS} = 1$ and $N_{RF}^{UE} = 1$. }
\label{fig_hbf}
\end{figure}

We further consider a system with subarray HAD-BF structures at both the BS and UE, depicted in Fig. \ref{fig_hbf}. In this setup, the received signal of a single subcarrier on an OFDM symbol can be expressed as
\begin{equation}
\label{eq1}
\mathbf{y}= \text{ }\mathbf{W}_{BB}^{BS}\mathbf{R}_{1}^{BS}\mathbf{W}_{RF}^{BS}\mathbf{R}_{2}^{BS}\mathbf{H_c}\mathbf{T}_{2}^{UE}\mathbf{V}_{RF}^{UE}\mathbf{T}_{1}^{UE}\mathbf{V}_{BB}^{UE}\mathbf{s}+\mathbf{n,}
\end{equation}
\noindent where $\mathbf{V}_{BB}^{UE}\in {{\mathbb{C}}^{N_{RF}^{UE}\times N_{s}}}$, $\mathbf{W}_{BB}^{BS}\in {{\mathbb{C}}^{{{N}_{s}}\times N_{RF}^{BS}}}$ denote the baseband (BB) digital beamforming matrix of the UE and BS. $\mathbf{T}_{1}^{UE}\in{{\mathbb{C}}^{N_{RF}^{UE}\times N_{RF}^{UE}}}$, $\mathbf{R}_{1}^{BS}\in {{\mathbb{C}}^{N_{RF}^{BS}\times N_{RF}^{BS}}}$ denote the mapped digital RF mismatch coefficients of the UE and BS. $\mathbf{V}_{RF}^{UE}\in {{\mathbb{C}}^{N_{RF}^{UE}\times N_{RF}^{UE}}}$, $\mathbf{W}_{RF}^{BS}\in {{\mathbb{C}}^{N_{RF}^{BS}\times N_{ant}^{BS}}}$ 
are the analog beamforming weights of the UE and BS. $\mathbf{T}_{2}^{UE}\in{{\mathbb{C}}^{N_{ant}^{UE}\times N_{ant}^{UE}}}$ and $\mathbf{R}_{2}^{BS}\in{{\mathbb{C}}^{N_{ant}^{BS}\times N_{ant}^{BS}}}$ denote the mapped analog RF mismatch coefficients. $\mathbf{H_c}\in {{\mathbb{C}}^{N_{ant}^{BS}\times N_{ant}^{UE}}}$ represents channel. $\mathbf{s}$ is the data streams. $\mathbf{n}$ is the complex white Gaussian noise. $N_{ant}^{UE}$ and $N_{ant}^{BS}$ are the number of antennas at the UE and BS.  $N_{RF}^{UE}$ and $N_{RF}^{BS}$  are the number of digital RF chains at the UE and BS. $N_s$ is the number of data streams. If we introduce the Kronecker product and use block diagonal structures of  $\mathbf{V}_{RF}^{UE}$ and $\mathbf{W}_{RF}^{BS}$, the equivalent signal model can be written as \cite{23} 
\begin{equation}
\label{eq2}
\begin{aligned}  
\mathbf{y}=\;& \underbrace{\mathbf{W}_{BB}^{BS}\mathbf{W}_{RF}^{BS}}_{{{\mathbf{W}}_{BS}}}\underbrace{\left( \mathbf{R}_{1}^{BS}\otimes {{\mathbf{I}}_{BS}} \right)\mathbf{R}_{2}^{BS}}_{{{\mathbf{R}}_{BS}}}\mathbf{H_c}\\
           & \underbrace{\mathbf{T}_{2}^{UE}\left( \mathbf{T}_{1}^{UE}\otimes {{\mathbf{I}}_{UE}} \right)}_{{{\mathbf{T}}_{UE}}}\underbrace{\mathbf{V}_{RF}^{UE}\mathbf{V}_{BB}^{UE}}_{{{\mathbf{V}}_{UE}}}\mathbf{s}+\mathbf{n},
\end{aligned}
\end{equation}
where $\mathbf{I}_{BS}$ and $\mathbf{I}_{UE}$ are identity matrices of size $N_{ant}^{BS}/N_{RF}^{BS}$ and $N_{ant}^{UE}/N_{RF}^{UE}$, respectively. ${\mathbf{V}}_{UE}$ and ${\mathbf{W}}_{BS}$ are formed by aggregating the digital and analog beamforming matrices, ${\mathbf{T}}_{UE}$ and ${\mathbf{R}}_{BS}$ subsume the digital and analog RF mismatch coefficients.

If the UE transmits $N_{RF}^{UE}$ orthogonal pilots, BS can estimate uplink channel as
\begin{equation}
\label{eq4}
{\mathbf{G}_{UE\to BS}}={{\mathbf{W}}_{BS}}\underbrace{{{\mathbf{R}}_{BS}}{{\mathbf{H_c}}}{{\mathbf{T}}_{UE}}}_{{\mathbf{G}^{'}}_{UE\to BS}}{{\mathbf{V}}_{UE}}+{{\mathbf{N}}_{BS}}.
\end{equation}
Similarly, the UE can estimate DL channel as
\begin{equation}
\label{eq5}
{\mathbf{H}_{BS\to UE}}={{\mathbf{W}}_{UE}}\underbrace{{{\mathbf{R}}_{UE}}{\mathbf{H_c}^{T}}{{\mathbf{T}}_{BS}}}_{{\mathbf{H}^{'}}_{BS\to UE}}{{\mathbf{V}}_{BS}}+{{\mathbf{N}}_{UE}}.
\end{equation}
where ${\mathbf{V}}_{BS}$ and ${\mathbf{W}}_{UE}$ are formed by aggregating digital and analog beamforming matrices of the BS and UE, ${\mathbf{T}}_{BS}$ and ${\mathbf{R}}_{UE}$ subsume digital and analog RF mismatch coefficients of the BS and UE. ${\mathbf{N}}_{BS}$ and ${\mathbf{N}}_{UE}$ are complex white Gaussian noise.

Traditionally, it is generally assumed that ${{\mathbf{H}}^{'}}_{BS\to UE}\ne {\mathbf{{G}^{'}}^T}_{UE\to BS}$, and 
therefore the RF mismatch coefficients on each antenna array need to be calibrated by traversing
${{\mathbf{W}}_{BS}}$, ${{\mathbf{V}}_{BS}}$, ${{\mathbf{W}}_{UE}}$ and ${{\mathbf{V}}_{UE}}$ to make them reciprocal\cite{23}.

In practical application scenarios based on HAD-BF architectures, the analog RF side at least guarantees a consistent directional map within the first dominant flap\cite{Yu2025}, so it can be assumed that
${{\mathbf{W}}_{BS}}=\mathbf{V}_{BS}^{T}$, ${{\mathbf{W}}_{UE}}=\mathbf{V}_{UE}^{T}$. Furthermore, for the precalibrated active antenna units (AAUs), it can be assumed that 
\begin{equation}\label{eq6}
\begin{aligned}
  {{\mathbf{R}}_{BS}}&={{\gamma }_{R}^{BS}}\mathbf{I}+{{\mathbf{R}}_{BS,\Delta }},\quad
  {{\mathbf{T}}_{BS}}={{\gamma }_{T}^{BS}}\mathbf{I}+{{\mathbf{T}}_{BS,\Delta }},\\
  {{\mathbf{T}}_{UE}}&={{\gamma }_{T}^{UE}}\mathbf{I}+{{\mathbf{T}}_{UE,\Delta }},\quad
  {{\mathbf{R}}_{UE}}={{\gamma }_{R}^{UE}}\mathbf{I}+{{\mathbf{R}}_{UE,\Delta }}.
\end{aligned}
\end{equation}
where, ${\gamma }_{R}^{BS}$,${\gamma }_{T}^{BS}$,${\gamma }_{T}^{UE}$,${\gamma }_{R}^{UE}$ are primary  amplitude-phase mismatches coefficients, ${\mathbf{R}}_{BS,\Delta }$,${\mathbf{T}}_{BS,\Delta }$,${\mathbf{T}}_{UE,\Delta }$,${\mathbf{R}}_{UE,\Delta }$ are the small residual mismatch coefficients of BS and UE. During channel estimation, each RF link completes one channel estimation, so we can simply assume the system has one RF link and one stream. Therefore, after substituting Eq.\ref{eq6} into Eq.\ref{eq4} and  Eq.\ref{eq5}, we have

\begin{align}
\label{eq6a}
  & {{{g}}_{UE\to BS}} = {{\gamma }_{R}^{BS}}{{\gamma }_{T}^{UE}}{{\mathbf{w}}_{BS}}{{\mathbf{H_c}}}{{\mathbf{v}}_{UE}} \notag \\ 
  & \phantom{{}={}} 
    \left. 
    \begin{aligned}
      &+ {{\gamma }_{R}^{BS}}{{\mathbf{w}}_{BS}}{{\mathbf{H_c}}}{{\mathbf{T}}_{UE,\Delta }}{{\mathbf{v}}_{UE}} \\
      &+ {{\gamma }_{T}^{UE}}{{\mathbf{w}}_{BS}}{{\mathbf{R}}_{BS,\Delta }}{{\mathbf{H_c}}^{T}}{{\mathbf{v}}_{UE}} \\
      &+ {{\mathbf{w}}_{BS}}{{\mathbf{R}}_{BS,\Delta }}{{\mathbf{H_c}}}{{\mathbf{T}}_{UE,\Delta }}{{\mathbf{v}}_{UE}} + {{\mathbf{n}}_{BS}}
    \end{aligned}
    \right\} \tilde{{n}}_{BS}, \tag{6a} 
\end{align}
\begin{align}  
\label{eq6b}
  & {{{h}}_{BS\to UE}} = {{\gamma }_{R}^{UE}}{{\gamma }_{T}^{BS}}{{\mathbf{w}}_{UE}}{{\mathbf{H_c}^{T}}}{{\mathbf{v}}_{BS}} \notag \\ 
  & \phantom{{}={}} 
    \left. 
    \begin{aligned}
      &+ {{\gamma }_{R}^{UE}}{{\mathbf{w}}_{UE}}{{\mathbf{H_c}^{T}}}{{\mathbf{T}}_{BS,\Delta }}{{\mathbf{v}}_{BS}} \\
      &+ {{\gamma }_{T}^{BS}}{{\mathbf{w}}_{UE}}{{\mathbf{R}}_{UE,\Delta }}{{\mathbf{C}}}{{\mathbf{v}}_{BS}} \\
      &+ {{\mathbf{w}}_{UE,\Delta}}{{\mathbf{R}}_{UE}}{{\mathbf{H_c}^{T}}}{{\mathbf{T}}_{BS,\Delta }}{{\mathbf{v}}_{BS}} + {{{n}}_{UE}}
    \end{aligned}
    \right\} \tilde{{n}}_{UE}. \tag{6b} 
\end{align}
where ${\mathbf{v}}_{UE}$ and ${\mathbf{w}}_{BS}$ and  can be seen as one column of $\mathbf{V}_{RF}^{UE}$ and $\mathbf{W}_{RF}^{BS}$. ${\mathbf{v}}_{BS}$ and ${\mathbf{w}}_{UE}$ and  can be seen as one column of $\mathbf{V}_{RF}^{BS}$ and $\mathbf{W}_{RF}^{UE}$. Thus, when  $\mathbf{R}_{BS,\Delta}$, $\mathbf{T}_{BS, \Delta}$,$\mathbf{T}_{UE, \Delta}$ and $\mathbf{R}_{U E, \Delta}$ equal to zero, then the calibration coefficient can be calculated as
\setcounter{equation}{6}
\begin{equation}
\begin{aligned}
c_{cal} =\frac{g_{UE \rightarrow BS}}{h_{BS \rightarrow UE}} =\frac{\gamma_{RT}^{ BS} \gamma_{T}^{UE}}{\gamma_{RT}^{UE} \gamma_{TT}^{ BS}}
\end{aligned}
\end{equation}

However, when  $\mathbf{R}_{BS,\Delta}$, $\mathbf{T}_{BS, \Delta}$,$\mathbf{T}_{UE, \Delta}$ and $\mathbf{R}_{U E, \Delta}$ not equal to zero, and assumed to be independently and identically distributed (i.i.d.) zero-mean Gaussian random variables, then, we have,
\begin{equation}
\begin{aligned}
\mathbb{E}[{c_{cal}}]=\frac{\gamma_{R}^{BS} \gamma_{T}^{UE}}{\gamma_{R}^{UE} \gamma_{T}^{BS}}
\end{aligned}
\end{equation}
The above expression holds under the assumption that ${\mathbf{w}_{BS}}{\mathbf{H_c}}{\mathbf{v}_{UE}}$ is a relatively strong signal component, significantly larger than the residual noise terms. This condition is easily met in practice. For HAD-BF systems, the first step is to perform beam training so that the beam gain between the transmit and receive beams is maximized.

Upon this simplification, only the channel coefficients require calibration to establish reciprocity between ${\mathbf{G}_{UE\to BS}}$ and ${\mathbf{H}_{BS\to UE}}$ in the context of the fixed analog beamforming. And the calibration coefficient can also be used for other analog beamforming as long as it satisfies the aforementioned assumption and Eq.\ref{eq6}. This approach entails lower calibration overhead than enforcing reciprocity between ${\mathbf{H}^{'}}_{BS\to UE}$ and ${\mathbf{G}^{'}}_{UE\to BS}$. In summary, to harvest the total-antenna gain of distributed MIMO, the task reduces to (i) pairwise beam-sweep alignment at every digital RF chain, and (ii) coherent collaboration across the distributed digital RF chain. It should be noted that, since the equivalent noises ${{\widetilde{{n}}}_{BS}}$ and ${{\widetilde{{n}}}_{UE}}$ are non-zero, the phases of calibration coefficients derived from different UE measurements may diverge under low SNR conditions. This behavior differs from systems employing omnidirectional antennas in lower frequency bands. This is also demonstrated by practical test results in Section \ref{coefcharacter}.

\subsection{Reciprocity Modeling with Time-Frequency Offsets in Multi-Point-to-Multi-Point Systems}
Without loss of generality, we consider a system with ${N}$ UEs and ${M}$ TRPs, each equipped with a single RF channel and fixed analog beamforming. The system collects the channel from ${K}$ subcarriers and ${L}$ OFDM symbols. The channel is commonly assumed sparse with only a few dominant paths in mmWave systems. Then, the ideal OTA channel of the ${n}$-th UE to the ${m}$-th TRP at the ${k}$-th subcarrier and the ${l}$-th symbol over coherence time can be modeled as\cite{Pucci2022}

\begin{equation}
\label{eq8}
\begin{aligned}
  \mathbf{\bar{H}}_{m,n}(k,l) &= \sum_{q=0}^{Q-1} \Big( \alpha_{mnq} \exp\left( -j 2\pi k \Delta f \frac{d_{mnq}}{c} \right) \\
  &\quad \cdot \exp\left( j 2\pi \frac{v_{mnq}}{c} f_c l T \right) \Big).
\end{aligned}
\end{equation}
where ${{\alpha }_{mnq}}$ denotes the complex gain of the ${q}$-th path from the ${n}$-th UE to the ${m}$-th TRP. $Q$ denotes the number of paths from the ${n}$-th UE to the ${m}$-th TRP. $\frac{{{d}_{mnq}}}{c}$ denotes the propagation delay of the ${q}$-th path from the ${n}$-th UE to the${m}$-th TRP. $\frac{{{v}_{mnq}}}{c}f_c$ denotes the Doppler of the ${q}$-th path from the ${n}$-th UE to the ${m}$-th TRP. $\Delta f$ denotes the subcarrier spacing of OFDM. ${{f}_c}$ denotes the carrier frequency. The information such as the distance and velocity of sensed scatterers of interest can be extracted from the OTA channel.

In a practical scenario, the actual uplink and downlink channel is distorted by the RF non-idealities discussed previously, along with oscillator-induced timing and frequency offsets with respect to the network timing. Thus, the uplink channel and the downlink channel can be modeled as \cite{Nissel2022}
\begin{subequations}
\label{eq11}
\small
\begin{align}
  \mathbf{G}_{m,n}(k,l) =\; & 
    \beta_{m}^{r}(k,l) \, \beta_{n}^{\mathrm{t}}(k,l) \mathbf{\bar{H}}_{m,n}(k,l) \nonumber \\
  &\cdot \exp\bigl(-j2\pi k \Delta f \tau_{n,m}\bigr) \exp\bigl(j2\pi e_{n,m} f_c l T\bigr), \label{eq11a} \\[1ex]
  \mathbf{H}_{m,n}(k,l) =\; & 
    \beta_{n}^{r}(k,l) \, \beta_{m}^{\mathrm{t}}(k,l) \mathbf{\bar{H}}_{m,n}(k,l)  \nonumber \\
  & \cdot \exp\bigl(j2\pi k \Delta f \tau_{n,m}\bigr)\exp\bigl(-j2\pi e_{n,m} f_c l T\bigr). \label{eq11b}
\end{align}
\end{subequations}
where $\beta _{m}^{t}\left( k,l \right)$ and $\beta _{m}^{r}\left( k,l \right)$ denote transmit and receive RF complex gain of the ${m}$-th TRP. $\beta _{n}^{t}\left( k,l \right)$ and $\beta _{n}^{r}\left( k,l \right)$ denote transmit and receive RF complex gain of the ${n}$-th UE. ${{e}_{n,m}}=e_{n}^{{}}-e_{m}^{{}}$ represents the frequency offset between the ${m}$-th TRP and the ${n}$-th UE in ppb. $e_{n}$ and $e_{m}$ denote the frequency offset of the UE and BS introduced by the crystal in ppb. ${{\tau }_{n,m}}=\tau _{n}^{{}}-\tau _{m}^{{}}$ represents the time offset between the ${m}$-th TRP and the ${n}$-th UE . $\tau_{n}$ and $\tau _{m}$ denote the time offset of the UE and BS with respect to the network genie time. 

If we further assume that the receiver channel complex gain remains the same throughout the bandwidth and does not vary with time, which is valid in static or quasi-static scenarios, then the ideal reciprocity calibration coefficient between the ${m}$-th TRP and the ${n}$-th UE is modeled as
\begin{equation}
\label{eq12}
\begin{aligned}
    &\mathbf{C}_{m,n}(k,l)= \frac{\mathbf{G}_{m,n}(k,l)}{\mathbf{H}_{m,n}(k,l)} \\
    &= \frac{\beta_{m}^{r}\beta_{n}^{\text{t}}}{\beta_{n}^{r}\beta_{m}^{\text{t}}} \exp \left( -j4\pi k\Delta f\tau_{n,m} \right) \exp \left( j4\pi e_{n,m}f_{c}lT \right).
\end{aligned}
\end{equation}

However, DL channel ${\mathbf{H}_{m,n}}\left( k,l \right)$ and UL channel ${\mathbf{G}_{m,n}}(k,{l}')$ are acquired at different time and the channel can be dynamic. The actual available reciprocity calibration coefficients are denoted by
\begin{equation}
\label{eq13}
\begin{aligned}   
    &\mathbf{C}_{m,n}(k,l') = \frac{\mathbf{G}_{m,n}(k,l)}{\mathbf{H}_{m,n}(k,l')}= \frac{\beta_{m}^{r} \beta_{n}^{\mathrm{t}}}{\beta_{n}^{r} \beta_{m}^{\mathrm{t}}} 
    \frac{\bar{\mathbf{H}}_{m,n}(k,l)}{\mathbf{\bar{H}}_{m,n}(k,l')} \\
    & \cdot \exp(-j4\pi k \Delta f \tau_{n,m}) \exp\bigl( j2\pi e_{n,m} f_c (l + l') T \bigr).
\end{aligned}
\end{equation}

As shown, the OTA channel cannot be cancelled in dynamic scenarios. Specifically, if the UE is in motion or a high-speed object is present in the environment, the reciprocity of the OTA channel is compromised by the Doppler effect. Therefore, it is necessary to integrate sensing with calibration in such scenarios, utilizing sensing reconstruction to restore channel reciprocity and obtain accurate calibration coefficients.

\subsection{Existing calibration algorithms}
\subsubsection{Argos Calibration Algorithm}
Traditional calibration algorithms typically require knowledge of both the uplink and downlink channels, denoted as $\mathbf{G}$ and $\mathbf{H}$, respectively. For simplicity, we omit the subcarrier and time slot indices in the following description. The Argos algorithm computes the calibration coefficients as
\begin{equation}
\mathbf{C}_{\text{Argos}} = \mathbf{G} \oslash \mathbf{H}^{T},
\end{equation}
where $\oslash$ denotes element-wise division. Each column of the resulting matrix $\mathbf{C}_{\text{Argos}}$ represents the calibration coefficients with respect to a specific user, using that user as the reference.

A refined version, referred to as Argos mean\cite{Cao2023}, averages the calibration coefficients across all users after special normalization. It is noteworthy that both Argos and Argos mean require the downlink channel $\mathbf{H}$ to be fed back to the BS, which then computes the calibration coefficients using the uplink channel $\mathbf{G}$ and the feedback channel $\mathbf{H}$.

\subsubsection{TLS Algorithm}
By exploiting both uplink and downlink channel measurements, the calibration problem can be formulated as a TLS optimization problem\cite{19}

\begin{equation}
\min_{\mathbf{C}_{BS},\mathbf{C}_{UE}}\left\|\mathbf{G}^{\mathrm{T}} \cdot \mathbf{C}_{BS}-\mathbf{C}_{UE} \cdot \mathbf{H}\right\|_{F}^{2},
\end{equation}
where $\mathbf{C}_{BS}$ and $\mathbf{C}_{UE}$ denote calibration coefficient of the BS and UE. Vectorizing the above expression yields

\begin{equation}
\begin{array}{l}
\operatorname{vec}\left(\mathbf{G}^{\mathrm{T}} \cdot \mathbf{C}_{BS}-\mathbf{C}_{U E} \cdot \mathbf{H}\right) \\
\quad=\left[\mathbf{I}_{M} \otimes \mathbf{G}^{\mathrm{T}},-\mathbf{H} \otimes \mathbf{I}_{N}\right] \mathbf{c}=\mathbf{Q c},
\end{array}
\end{equation}
where $\otimes$ denotes the Khatri-Rao product, and
\begin{equation}
\mathbf{c} \doteq \left[ \left( \widetilde{\operatorname{diag}} \left\{ \mathbf{C}_{A} \right\} \right)^{\mathrm{T}}, \left( \widetilde{\operatorname{diag}} \left\{ \mathbf{C}_{B} \right\} \right)^{\mathrm{T}} \right]^{\mathrm{T}},
\end{equation}
\begin{equation}
\mathbf{Q} \doteq \left[\mathbf{I}_{M} \otimes \mathbf{G}^{\mathrm{T}},-\mathbf{H} \otimes \mathbf{I}_{N}\right].
\end{equation}

Therefore, the problem can be reformulated as,
\begin{equation}
\min _{c} \mathbf{c}^{\mathrm{H}} \mathbf{Q}^{\mathrm{H}} \mathbf{Q c} .
\end{equation}
By introducing a norm constraint $\| \mathbf{c} \|_2^2 = 1$ to avoid the trivial solution, the optimal $\mathbf{c}$ is given by the eigenvector corresponding to the smallest eigenvalue of the matrix $\mathbf{Q}^\mathsf{H}\mathbf{Q}$.

\subsubsection{Other Variants and Enhancements}
Most existing calibration algorithms are variants or extensions of the Argos and TLS methods. Recently, neural network-based approaches have been proposed to leverage their powerful feature extraction capabilities\cite{Xu2024}. These methods enhance the estimation of uplink and downlink channels, effectively reducing noise and improving the accuracy of calibration coefficients.

Additionally, graph-theoretic approaches such as spanning tree algorithms\cite{17} have been employed to optimize the grouping of multiple nodes in distributed systems. By identifying the optimal tree topology, these methods apply calibration algorithms (e.g., Argos) in an indirect manner, which has been shown to achieve higher SNR compared to direct calibration processes.

\section{Bidirectional Calibration and Synchronization}
\label{section3}
\subsection{Bidirectional Calibration Process Design in Quasi-Static Scenarios}
\begin{figure}[htbp]
\centering
\includegraphics[width=0.75\linewidth]{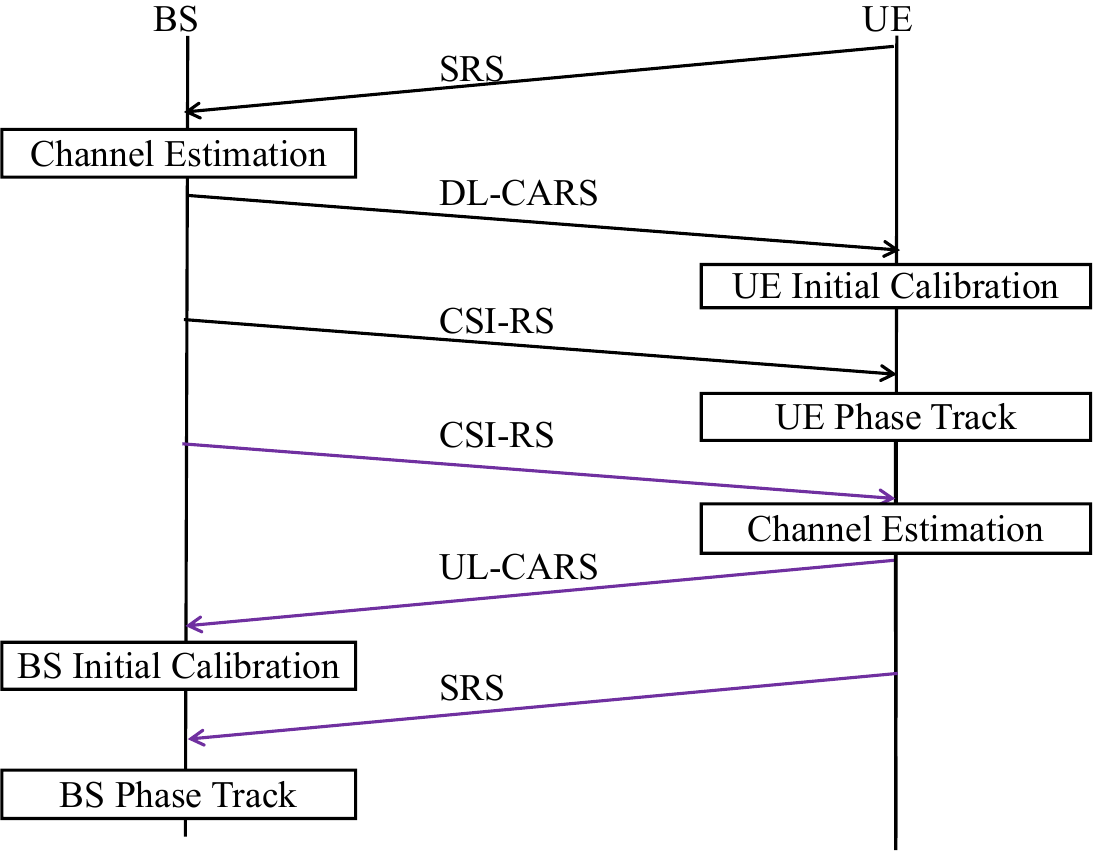}
\caption{The bidirectional calibration process in a quasi-static scenario.}
\label{fig_staticprocedure}
\end{figure}

As is shown in Fig. \ref{fig_staticprocedure}, the bidirectional calibration flow in the quasi-static scenario is designed as follows. The UE side transmits sounding reference signal (SRS) to the BS at the $l$-th OFDM symbol, and the BS transmits the conjugate precoded DL calibration reference signal (DL-CARS) at the  ${l}'$-th OFDM after the BS completes the channel estimation. Then, the UE completes the initial RF calibration coefficients estimation of the UE side by using the DL-CARS. In the subsequent communication process, the BS can periodically transmit the DL channel state information reference signal (CSI-RS) at the $l''$-th OFDM, and the UE utilizes the CSI-RS to complete the phase tracking of the calibration coefficients, thereby reducing the calibration overhead.

Similarly, the BS side transmits CSI-RS to the UE at the  $l$-th OFDM symbol, and the UE transmits a conjugate precoded UL calibration computation reference signal (UL-CARS) at the  $l'$-th OFDM after the UE has completed the channel estimation. Then, the BS completes the initial RF calibration coefficients estimation at the BS side using the UL-CARS. In the subsequent communication process, the UE periodically transmits the UL SRS pilot at the  $l''$-th OFDM, and the BS completes the phase tracking of the calibration coefficients by utilizing the UL SRS, thus reducing the calibration overhead.

In order to realize the above auxiliary calibration, a feasible orthogonal pilot mapping method is given as follows, assuming that there are  ${M}$ TRPs at the BS side, the frequency domain is combed with  ${M}$.  After the BS transmits the M subcarrier signals to the UE, the UE further divides the subcarriers based on the number of assisted calibration UEs $N$ in the system. This ensures that the signals transmitted from the UE to the BS are orthogonal and do not interfere with each other. It is expressed as follows by Eq. \ref{eq14}.
\begin{subequations}
\label{eq14}
\begin{align}
k'_{bs} &= kM + n_{bs}^{\mathrm{comb}}, 0 \le k \le K-1, \label{eq14a} \\
k'_{ue} &= kMN + n_{ue}^{\mathrm{comb}} M +n_{bs}^{\mathrm{comb}},  0 \le k \le \frac{K}{N}-1, \label{eq14b}
\end{align}
\end{subequations}
 where ${{{k}'}_{bs}}$ denotes the frequency domain mapping index of the BS side, ${{{k}'}_{ue}}$ denotes the frequency domain mapping index of the UE side, ${{n}_{bs}^{comb}}$ denotes the corresponding BS antenna port, and ${{n}_{ue}^{comb}}$ denotes the corresponding UE antenna port. $1\le n_{bs}^{\mathrm{comb}}\le M$,$1\le n_{bs}^{\mathrm{comb}}\le N$. $K$ indicates the length of the pilot base sequence length on the BS side.

\emph{Remark:} We only provide one feasible frequency-division orthogonal pilot mapping method here, and it can be further expanded to use multiple consecutive OFDMs or multiple orthogonal codes to generate more orthogonal pilots in order to support more UE and BS antennas. 

\begin{figure}[htbp]
\centering
\includegraphics[width=0.75\linewidth]{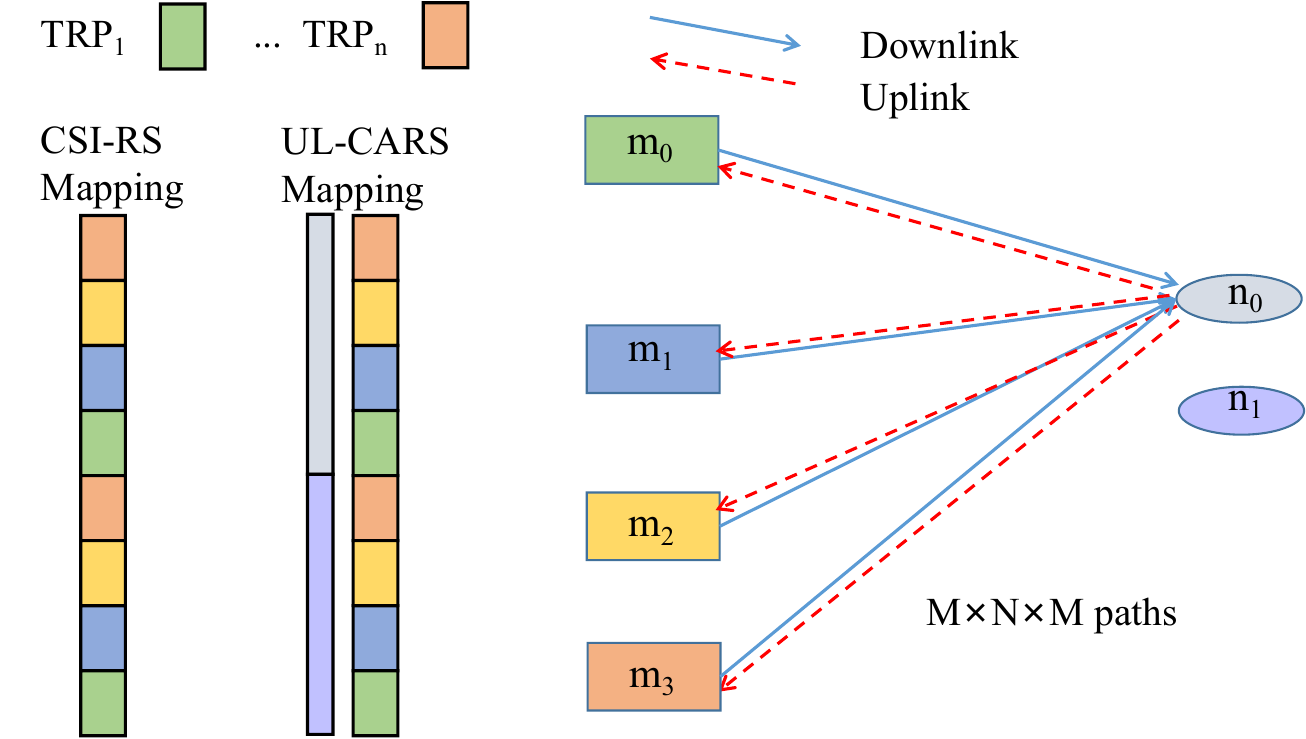}
\caption{Pilot mapping method and transmission diagram.}
\label{fig_ceMap}
\end{figure}

\subsection{Initial RF Calibration Coefficient Estimation}
Firstly, the BS transmits  ${M}$ orthogonal DL CSI-RS reference signals to  ${N}$ UEs at the ${l}$-th OFDM symbol, and the UEs obtain the DL channel  ${\mathbf{H}}$. After the UEs obtain the channel, the UEs use conjugate precoding at the  ${l'}$-th OFDM symbol to generate the CARS, each CARS contains ${M}$ orthogonal pilots. After the CARS are transmitted to the BS, the equivalent UL channel obtained on the BS side is
\begin{equation}
\label{eq15}
\begin{aligned}
\mathbf{\widehat{G}}_{m,n,m'}(k, l') 
&= \frac{\beta_m^{\mathrm{r}} \beta_n^{\mathrm{t}}}{\beta_{m'}^{\mathrm{t}} \beta_n^{\mathrm{r}}} 
\cdot \frac{\mathbf{\bar{H}}_{n,m}(k,l)}{\mathbf{\bar{H}}_{n,m'}(k,l')} \\
&\quad \cdot \exp\bigl(-j 2\pi k \Delta f (\tau_{n,m} + \tau_{n,m'})\bigr) \\
&\quad \cdot \exp\bigl(j 2\pi f_c T (e_{n,m} l + e_{n,m'} l')\bigr),
\end{aligned}
\end{equation}
where $m'$ indicates the transmitting BS antenna port, $m$ indicates the receiving BS antenna port, $n$ indicates the UE antenna port.
For simplicity, only the equivalent channel at $m' = m$  is estimated, that is
\begin{equation}\label{eq16}
\begin{aligned}
\mathbf{\widehat{G}}_{m,n}(k,l') \approx\ 
& \frac{\beta_m^{\mathrm{r}} \beta_n^{\mathrm{t}}}{\beta_m^{\mathrm{t}} \beta_n^{\mathrm{r}}}
\cdot \exp\left(-j 4\pi k \Delta f\, \tau_{n,m} \right) \\
& \cdot \exp\left(j 4\pi e_{n,m} l'\, f_c T \right).
\end{aligned}
\end{equation}
Here, we assume the ${{\tau }_{n,m}}$ and ${{e}_{n,m}}$ within the  ${l}$-th and  ${l'}$-th OFDM symbols are similar. Similarly, the estimated equivalent channel estimation can be obtained at the UE side
\begin{equation}\label{eq17}
\begin{aligned}
\mathbf{\widehat{H}}_{m,n}(k,l') \approx\ 
& \frac{\beta_n^{\mathrm{r}} \beta_m^{\mathrm{t}}}{\beta_n^{\mathrm{t}} \beta_m^{\mathrm{r}}}
\cdot \exp\left(j 4\pi k \Delta f\, \tau_{n,m} \right) \\
& \cdot \exp\left(-j 4\pi e_{n,m} l' f_c T \right).
\end{aligned}
\end{equation}

It is clear to find that the equivalent channel obtained at the BS side and the equivalent channel obtained at the UE side are formally the same. The algorithm for solving the calibration coefficients is given below as an example at the BS side. The equivalent channel obtained at the UE side can be solved to obtain the calibration coefficients according to the same steps.

We propose a two-step maximum likelihood (ML) TLS parameter estimation method to estimate the delay and the initial calibration coefficients from the BS side. For the initial calibration estimation, the phase term $\exp \left( j4\pi {{e}_{n,m}}{{l}^{'}}{{f}_{c}}T \right)$ can be regarded as one, and the equivalent estimated channel model can be rewritten as the classical parameter estimation model as
\begin{equation}
\label{eq18}
{{\mathbf{g}}_{m,n}}={{c}_{n,m}}{\mathbf{\varphi}}({{\tau }_{n,m}})+{{\mathbf{n}}_{m,n}}.
\end{equation}
where  ${\mathbf{\varphi}}({{\tau }_{n,m}})={{[1,\ldots ,\exp (-j4\pi (K-1)\Delta f{{\tau }_{n,m}})]}^{T}}$, ${{c}_{n,m}}=\frac{\beta_{m}^{\text{r}}\beta_{n}^{\text{t}}}{\beta_{m}^{\text{t}}\beta _{n}^{r}}$, ${{\mathbf{n}}_{m,n}}$ is a complex Gaussian noise vector.

In the first step, the delay and initial relative calibration coefficient estimates are obtained using the ML estimation. Maximizing the likelihood function is equivalent to minimizing the residual term in the exponential,
\begin{equation}\label{eq19}
  \mathcal{L}({\hat{c}_{n,m}},{\hat{\tau }_{n,m}})=\min \|{{\mathbf{g}}_{m,n}}-{\hat{c}_{n,m}}{\mathbf{\varphi}}({\hat{\tau }_{n,m}}){{\|}^{2}},
\end{equation}
where, $\hat{c}_{n,m}$ is the estimated complex RF coefficient between the ${n}$-th UE and the ${m}$-th TRP. $\hat{\tau}_{n,m}$ is the estimated timing offset between the ${n}$-th UE and the ${m}$-th TRP.
Therefore, the ML estimation is
\begin{subequations}\label{eq19m}
\begin{align}
  \hat{\tau}_{n,m} &= \arg \max_{\boldsymbol{\tau}_{n,m}} 
  \left| \boldsymbol{\varphi}^H({\tau}_{n,m})\, \mathbf{g}_{m,n} \right|^2, \label{eq19a} \\[0.5ex]
  \widehat{c}_{n,m} &= \frac{\boldsymbol{\varphi}^H(\hat{\tau}_{n,m})\, \mathbf{g}_{m,n}}
  {\boldsymbol{\varphi}^H(\hat{\tau}_{n,m})\, \boldsymbol{\varphi}(\hat{\tau}_{n,m})}.\label{eq19b}
\end{align}
\end{subequations}

In the second step, the TLS method is used to further estimate the RF calibration coefficients with $M\times N$ ML estimation $\widehat{c}_{n,m}$.
The maximum likelihood estimation of ${M}$ nodes at the BS side, which can be further expressed in terms of the RF calibration coefficients as
\begin{equation}
\label{20}
  {{\widehat{c}}_{n,m}}=\frac{\beta _{m}^{\text{r}}\beta _{n}^{\text{t}}}{\beta _{n}^{r}\beta _{m}^{\text{t}}}=\frac{{{\widetilde{c}}_{\text{UE,}n}}}{{{\widetilde{c}}_{\text{BS,}m}}}.
\end{equation}
where ${{\widetilde{c}}_{UE,n}}$ is the initial RF calibration coefficient of the $n$-th node on the UE side and ${{\widetilde{c}}_{BS,m}}$ is the initial RF calibration coefficient of the $m$-th node on the base station side. The algorithm for the joint calibration of the BS antenna ports and the UE antenna ports using the TLS method can be described as
\begin{equation}
\label{21}
\begin{aligned}
  \min & \; J\big(\widetilde{c}_{\mathrm{BS},1}, \widetilde{c}_{\mathrm{BS},2}, \cdots, \widetilde{c}_{\mathrm{UE},1}, \widetilde{c}_{\mathrm{UE},2}, \cdots \big) \\
       = & \sum_{n=1}^{N} \sum_{m=1}^{M} \left| \widetilde{c}_{\mathrm{BS},m} \widehat{c}_{n,m} - \widetilde{c}_{\mathrm{UE},n} \right|^2
\end{aligned}
\end{equation}
where, $\mathbf{c} = [\widetilde{c}_{\mathrm{BS},1}, \widetilde{c}_{\mathrm{BS},2}, \cdots, \widetilde{c}_{\mathrm{UE},1}, \widetilde{c}_{\mathrm{UE},2}, \cdots \big]$.

It can be written in the form of ${\mathbf{c}^{\text{H}}}\mathbf{A}\mathbf{c}$, where
\begin{subequations}
\label{eq_chc}
\begin{equation}
\mathbf{A} = \begin{bmatrix}
\mathbf{A}_{11} & \mathbf{A}_{12} \\
\mathbf{A}_{21} & \mathbf{A}_{22}
\end{bmatrix},
\end{equation}
\begin{equation}
\mathbf{A}_{11} = \text{diag} \left( \sum_{n=1}^{N} |c_{n,1}|^2, \sum_{n=1}^{N} |c_{n,2}|^2, \dots, \sum_{n=1}^{N} |c_{n,M}|^2 \right),
\end{equation}
\begin{equation}
\mathbf{A}_{21} = \begin{bmatrix}
-\hat{c}_{1,1} & -\hat{c}_{2,1} & \cdots & -\hat{c}_{N,1} \\
-\hat{c}_{1,2} & -\hat{c}_{2,2} & \cdots & -\hat{c}_{N,2} \\
\vdots & \vdots & \ddots & \vdots \\
-\hat{c}_{1,M} & -\hat{c}_{2,M} & \cdots & -\hat{c}_{N,M}
\end{bmatrix},
\end{equation}
\begin{equation}
\mathbf{A}_{12} = \mathbf{A}_{21}^{H},\quad \mathbf{A}_{22} = M \cdot \text{diag}(1, 1, \dots, 1)_{M},
\end{equation}
\end{subequations}

The solution sought for $\mathbf{c}$ is the eigenvector corresponding to the smallest eigenvalue of the matrix $\mathbf{A}$,
\begin{equation}
\label{tlssolve}
A \mathbf{c} = \lambda_{\text{min}} \mathbf{c},
\end{equation}
where $\lambda_{\text{min}}$ represents the smallest eigenvalue of matrix $\mathbf{A}$, and $\mathbf{c}$ is the corresponding eigenvector.

\subsection{Phase Tracking of Calibration Coefficients}
After estimating the initial calibration coefficients, a reciprocal calibration coefficient tracking scheme based on the UL SRS signal is further proposed.

From Eq. \ref{eq16}, after the initial calibration coefficients are determined, the subsequent calibration coefficients are actually changed in phase over time. According to the previous modeling, the ideal calibration coefficients of the $l'$-th and  $l''$-th OFDMs can be expressed as
\begin{equation}
\label{eq23}
  {{\mathbf{C}}_{m,n}}\left( k,{l}'' \right)={\mathbf{C}_{m,n}}\left( k,{l}' \right)\exp \left[ j4\pi {{e}_{n,m}}{{f}_{c}}\left( {l}''-{l}' \right)T \right].
\end{equation}
Therefore, after obtaining the calibration coefficients of the  ${l'}$-th OFDM symbol, the calibration coefficients of the  ${l''}$-th OFDM can be obtained by phase tracking. From Eq. \ref{eq11}, the channel obtained by the  ${l'}$-th and  ${l''}$-th OFDMs through the UE UL SRS can be expressed as follows
\begin{subequations}
\label{eq24}
\begin{align}
\mathbf{G}_{m,n}(k,l') &= \beta_{m,n} \exp\left(j2\pi k \Delta f \tau_{n,m}\right) \notag \\
&\quad \times \exp\left(-j2\pi e_{n,m} f_c l' T\right) \mathbf{\bar{H}}_{n,m}(k,l'), \label{eq24a} \\
\mathbf{G}_{m,n}(k,l'') &= \beta_{m,n} \exp\left(j2\pi k \Delta f \tau_{n,m}\right) \notag \\
&\quad \times \exp\left(-j2\pi e_{n,m} f_c l'' T\right) \mathbf{\bar{H}}_{n,m}(k,l''), \label{eq24b}
\end{align}
\end{subequations}

In the quasi-static scenario, the inserted UL SRS density is larger, then the OTA channels of the  ${l'}$-th and  ${l''}$-th OFDMs are similar and the OTA channels can be canceled. Then the phase tracking of the calibration coefficients can be realized by simply tracking the phase of the UE's UL SRS,
\begin{equation}
\label{eq_phasetrack}
\begin{split}
  \widehat{\phi}_{m,n}(k, l'') &= \text{angle}\left( \frac{G_{m,n}(k, l'')}{G_{m,n}(k, l')} \right) \\
  &= -j 2\pi e_{n,m} f_c \left( l'' - l' \right) T.
\end{split}
\end{equation}
Therefore, the calibration coefficients of the  ${l''}$-th OFDM can be expressed using phase tracking as
\begin{equation}
\label{25}
  {\mathbf{C}_{m,n}}\left( k,{l}'' \right)={\mathbf{C}_{m,n}}\left( k,{l}' \right)\exp \left(2{{\widehat{\phi }}_{m,n}}(k,{l}'')\right).
\end{equation}

Similarly, the calibration coefficients calculated at the UE side can be realized by simply tracking the DL CSI-RS.

\subsection{Simulation Results}
\begingroup 
\begin{table}[htbp]
  \centering
  \caption{Simulation Parameters}
  \label{tab:sim_param}
  \begin{tabular}{lc}
    \toprule
    \textbf{Parameter} & \textbf{Value} \\
    \midrule
    Carrier frequency $f_c$& 26 GHz \\
    Subcarrier spacing $\Delta f$& 120 kHz \\
    Time intervel $T$ & 0.625e-3 ms ( $\approx$ 70 OFDM)\\
    FFT size & 2048 \\
    System bandwidth & 200 MHz \\
    CP length & 144 \\
    Noise power spectral density & $-174$ dBm/Hz \\
    Number of BS AAUs & 8 \\
    Number of UE AAUs & 8 \\
    Number of antennas per AAU & 16 \\
    Channel model & Free-space path-loss model \\
    RF gains amplitude $|\beta|$  & $\mathcal{N}(0,0.1)$ \\
    RF gains phase $\angle(\beta)$ & $\mathcal{U}(-\pi, \pi)$ \\
    Time offset $\tau$&  $\mathcal{U}(-10, 10)$  ns \\
    Frequency offset $e$ &  $\mathcal{U}(-30, 30)$ ppb \\
    \bottomrule
  \end{tabular}
\end{table}
\endgroup
We conducted simulations under the fundamental parameters listed in Table \ref{tab:sim_param}, where the OTA channel is modeled by the free-space path-loss model. The timing offset and frequency offset of every AAU are assumed to be independent and uniformly distributed. Their peak values are set to 10 ns and 30 ppb respectively, representing the state-of-the-art synchronization accuracy currently achievable in commercial networks. We use 4 AAUs to calibrate 8 BS AAUs, and the remained 4 AAUs to calculate spectral efficiency.

\begin{figure}[htbp]
\centering
\includegraphics[width=0.65\linewidth]{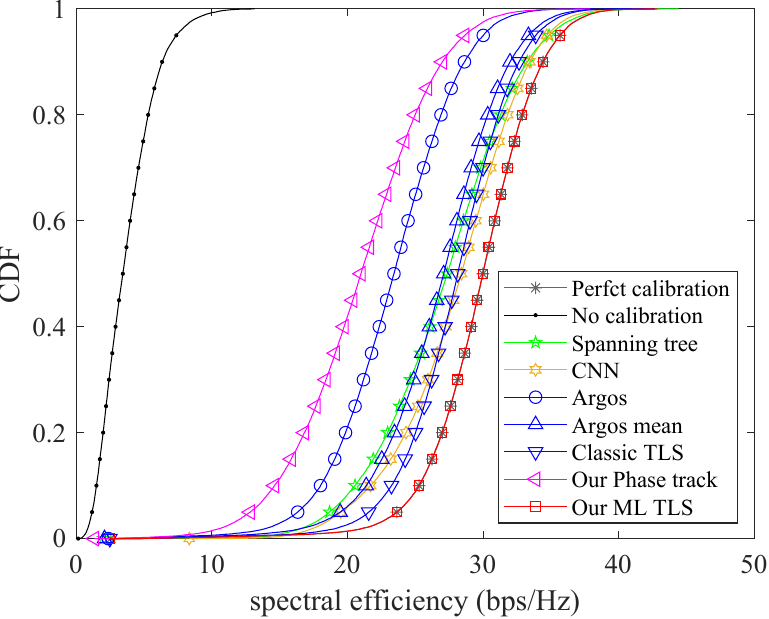}
\caption{Quasi-static scene spectral efficiency result for different algorithms with the transmission power of 30 dBm.}
\label{fig_staticCDF}
\end{figure}

Fig. \ref{fig_staticCDF} demonstrates that the proposed two-step ML TLS outperforms the traditional Argos, Argos mean, TLS, spanning tree and CNN algorithms. Thanks to the ML estimator, we effectively boost the SNR by exploiting multiple sub-carriers, thereby suppressing phase noise in the calibration coefficients. Besides, the TLS algorithm capitalizes on information across multiple transmit–receive pairs, yielding spatial diversity gain. When low-complexity phase tracking is implemented, the algorithm’s performance becomes comparable to that of the real-time Argos algorithm.

\section{Integrated calibration and sensing in dynamic scenarios}
\label{section4}
\subsection{Bidirectional Calibration Process Design in Dynamic Scenarios}

In scenarios where the UE is in dynamic motion or in the presence of moving scatterers, the calibration coefficients phase tracking algorithm applicable to quasi-static scenarios will have a serious performance loss because the reciprocity of the OTA channel does not exist. Therefore, in order to realize the calibration coefficient phase tracking in dynamic scenarios, it is necessary to utilize sensing to eliminate the phase variations of OTA channel to achieve accurate calibration coefficient phase tracking.

\begin{figure}[t]
\centering
\includegraphics[width=0.75\linewidth]{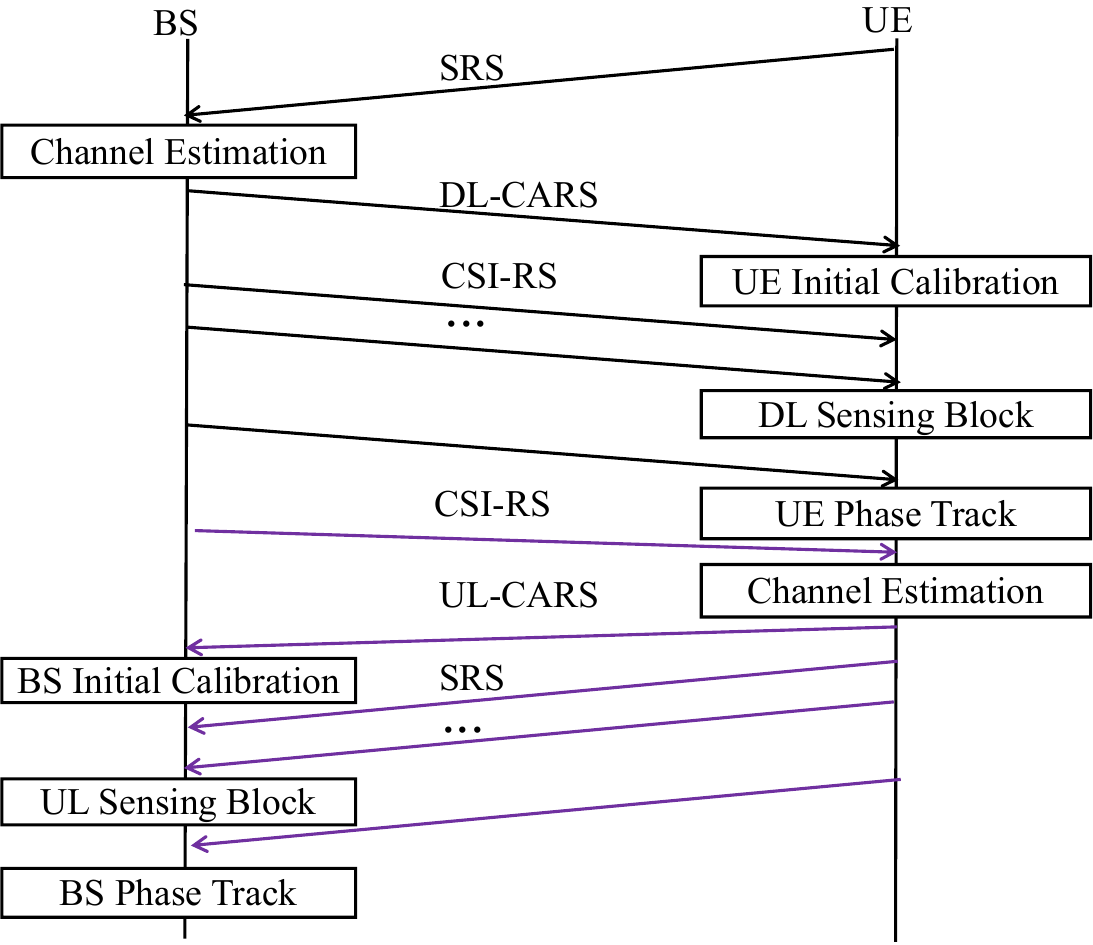}
\caption{The bidirectional calibration process in a dynamic scenario.}
\label{fig_moveprocedure}
\end{figure}

As is shown in Fig \ref{fig_moveprocedure}, the UE transmits SRS at the  $l$-th OFDM symbol UE side to the BS, and the BS transmits the conjugate precoded DL-CARS at the $l'$-th OFDM after completing channel estimation. Then the BS subsequently transmits the DL CSI-RS in succession with a short period. The UE completes the initial calibration coefficients estimation by utilizing the DL-CARS. The UE utilizes the DL CSI-RS transmitted continuously within short periods and initial calibration coefficents to complete the sensing of the OTA channel. In the subsequent communication process, the BS periodically transmits the DL CSI-RS at the $l''$-th OFDM, and the UE completes the phase tracking of the calibration coefficients by utilizing the DL CSI-RS and the sensed OTA channel, thus reducing the calibration overhead. 


Similarly, the BS side performs a symmetric operation in coordination with the UE side to complete the calibration and sensing. The following takes the UE transmitting the UL-CARS as an example, and gives the specific flow of realizing calibration-assisted sensing and sensing-assisted calibration at the BS side.

\subsection{Calibration-Assisted Sensing}
\label{cas}
If the UL SRS of the UE is directly utilized for sensing, the sensed OTA channel is affected by non-ideal factors, i.e., the interference of timing and frequency offset between transmitting and receiving nodes, and therefore reduces the time and speed sensing accuracy of target sensing. Therefore, it is firstly necessary to utilize the accurate estimation of the calibration coefficients over a period of time from the initial calibration. So, we have

\begin{equation}
\label{eq26}
\begin{aligned}
{\mathbf{C}_{n,m}}(k,l)&=\frac{\beta _{m}^{\text{r}}\beta _{n}^{\text{t}}}{\beta _{n}^{r}\beta _{m}^{\text{t}}}\exp \left( -j4\pi k\Delta f{{\tau }_{n,m}} \right)\\
 &\cdot \exp \left( j4\pi {{e}_{n,m}}.{{f}_{c}}lT \right)
\end{aligned}
\end{equation}

By utilizing the calibration coefficients, the effects of phase changes due to timing and frequency offset can be eliminated, and the residual RF complex coefficients do not affect the estimation of the sensed channel.

Therefore, the OTA sensing channel with the elimination of non-ideal factors can be recovered as
\begin{align}
\label{eq27}
  & {{\widehat{\mathbf{\bar{H}}}}_{n,m}}(k,l)={\mathbf{G}_{m,n}}(k,l)\sqrt{{\mathbf{{C}}_{n,m}}{{(k,l)}^{*}}} \notag\\ 
 & =\beta _{m}^{r}\beta _{n}^{\text{t}}\sqrt{{{\left( \frac{\beta _{m}^{\text{r}}\beta _{n}^{\text{t}}}{\beta _{n}^{r}\beta _{m}^{\text{t}}} \right)}^{*}}}{{\mathbf{\bar{H}}}_{n,m}}(k,l).
\end{align}

where the complex coefficients $\beta_{m}^{r}\beta_{n}^{\text{t}}\sqrt{{{\left(\frac{\beta_{m}^{\text{r}}\beta_{n}^{\text{t}}}{\beta_{n}^{r}\beta_{m}^{\text{t}}}\right)}^{*}}}$ are considered constant for a short period of time and do not affect the utilization of the channel to complete the sensing of velocity and distance. The channel sensing parameters can be extracted from estimed and calibrated ${{\widehat{\mathbf{\bar{H}}}}_{n,m}}(k,l)$ using two-dimensional fast fourier transform (2D-FFT) \cite{Sturm2011} and constant false alarm rate (CFAR) detection  \cite{Miller2009} based algorithms.

\subsection{Sensing-Assisted Calibration}
In a dynamic channel scenario, if the inserted UL SRS density is small, the OTA channels of the ${l'}$-th and ${l''}$-th OFDM are not similar. The phase difference between the calibration coefficient phase and the UL SRS channel phase is influenced by a dynamic interference phase that is related to the OTA channel. Therefore, calibration coefficient phase tracking based on OTA reciprocity will fail. 

Observing the UL SRS channel phase, it is easy to realize that the UL SRS channel phase contains not only the phase change of the calibration coefficients, but also the phase change of the OTA channel, which is shown in

\begin{equation}
\label{eq28}
\text{angle}\left( \frac{{\mathbf{G}_{m,n}}(k,{l}'')}{{\mathbf{G}_{m,n}}(k,{l}')} \right)={{\widehat{\phi }}_{m,n}}(k,{l}'')+\text{angle}\left( \frac{{{\mathbf{\bar{H}}}_{n,m}}(k,{l}'')}{{{\mathbf{\bar{H}}}_{n,m}}(k,{l}')} \right).
\end{equation}
where ${{\widehat{\phi }}_{m,n}}(k,{l}'')$ is the desired calibration phase change between the $l'$ and $l''$-th OFDM defined in Eq.\ref{eq_phasetrack}.  

After acquiring the sensing parameters of the calibrated OTA channel in section \ref{cas}, they can be further used to assist the calibration. In the dynamic scene, the  OTA channel at the next time step can be precidated using the sensed channel from the previous time step as
\begin{equation}
\label{eq29}
\begin{split}
\mathbf{\bar{H}}_{n,m}(k,l) = &\sum_{q=0}^{Q-1} \widehat{\alpha}_{mnq} 
\exp\left(-j2\pi k \Delta f \frac{\widehat{d}_{mnq}}{c}\right) \\
& \cdot \exp\left(j2\pi \frac{\widehat{v}_{mnq}}{c} f_c l T \right).
\end{split}
\end{equation}
where ${\alpha}_{mnq}$, ${\widehat{d}_{mnq}}$ and ${\widehat{v}_{mnq}}$ is the estimation of  the calibrated OTA channel.

The dynamic interference phase can be eliminated by utilizing the sensed reconstructed OTA channel, as shown in the following Eq. \ref{eq30}. This method can reduce the calibration insertion requirement and realize the phase tracking of the calibration coefficients under the dynamic scenarios.
\begin{equation}
\label{eq30}
  {{\widehat{\phi }}_{m,n}}(k,{l}'')=\text{angle}\left( \frac{{\mathbf{G}_{m,n}}(k,{l}'')}{{\mathbf{G}_{m,n}}(k,{l}')} \right)-\text{angle}\left( \frac{{{\mathbf{\bar{H}}}_{n,m}}(k,{l}'')}{{{\mathbf{\bar{H}}}_{n,m}}(k,{l}')} \right).
\end{equation}

Therefore, the ideal calibration coefficient for the  ${l''}$-th OFDM can be expressed using phase tracking as
\begin{equation}\label{eq31}
  {{\mathbf{C}}_{m,n}}\left( k,{l}'' \right)={\mathbf{C}_{m,n}}\left( k,{l}' \right)\exp \left({2{\widehat{\phi }}_{m,n}}(k,{l}'')\right).
\end{equation}

However, in practical scenarios, calibration is inevitably affected by the error of sensing channel reconstruction. For simplicity, taking a single path as an example, the Cramér–Rao lower bound (CRLB) of distance estimation and velocity estimation\cite{Sakhnini2021} is

\begin{subequations}
\label{eq:crlb}
\begin{align}
d_{\mathrm{CRLB}} &= \frac{3}{2 \rho \pi^{2}\left(\frac{\Delta_{\mathrm{f}}}{c}\right)^{2} K\left(K^{2}-1\right) L}, \\
v_{\mathrm{CRLB}} &= \frac{3}{2 \rho \pi^{2}\left(\frac{T}{\lambda}\right)^{2} L\left(L^{2}-1\right) K},
\end{align}
\end{subequations}

When predicting the channel, the maximum phase difference is achieved at the $K,L$-th element according to the channel formula. Thus, the phase error variance caused by distance estimation and velocity estimation\cite{Sakhnini2021} is

\begin{subequations}
\label{eq:phase}
\begin{align}
{\theta_{d}^{C R L B}=\frac{6 K^{2}}{\rho K L\left(K^{2}-1\right)}}, \\
 {\theta_{V}^{C R L B}=\frac{6 L^{2}}{\rho K L\left(L^{2}-1\right)}},
\end{align}
\end{subequations}

Consequently, the phase estimation error variance $\theta$ of the OTA channel is,

\begin{equation}
{\theta \geq \frac{6 K^{2}}{\rho K L\left(K^{2}-1\right)}+\frac{6 L^{2}}{\rho K L\left(L^{2}-1\right)}}
\label{eq:phase_all}
\end{equation}

It can be observed that the actual calibration coefficient will be affected by the phase estimation noise of channel reconstruction, and is related to the SNR, the number of subcarriers, and the number of OFDM symbols.. Additionally, the calibration coefficient is also influenced by channel estimation noise.

\begin{figure}[htbp]
\centering
\includegraphics[width=0.65\linewidth]{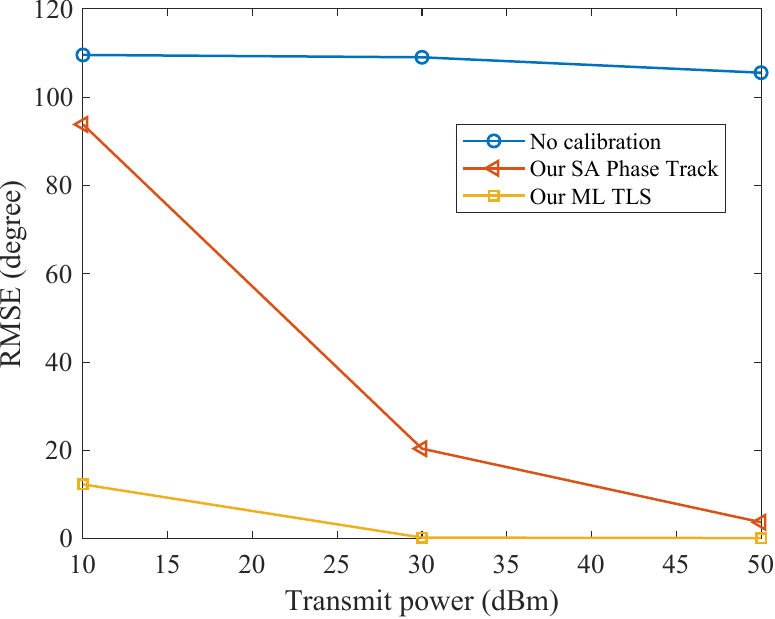}
\caption{Phase RMSE of the calibration coefficient versus transmit power.}
\label{fig:phaseRMSE}
\end{figure}

Fig.~\ref{fig:phaseRMSE} simulated the phase root mean square error (RMSE) for different transmit powers. 
Without phase tracking, the calibration phase error remains above $90^{\circ}$ regardless of the power level, rendering calibration ineffective in dynamic scenarios. 
By contrast, the proposed ML-TLS algorithm constrains the error to below $20^{\circ}$ even at $10\,\text{dBm}$. 
The tracking-based calibration scheme exhibits a clear power-dependent behavior. At $30\,\text{dBm}$ the phase error is approximately $20^{\circ}$, sufficient for multi-port coordination, whereas the error surges back to $90^{\circ}$ when the power drops below $10\,\text{dBm}$. 
This degradation stems from the propagation of both channel-reconstruction and channel-estimation errors into the calibration coefficient computation under low-SNR conditions.

\subsection{Simulation Results}
\begin{figure}[t]
\centering
\includegraphics[width=0.65\linewidth]{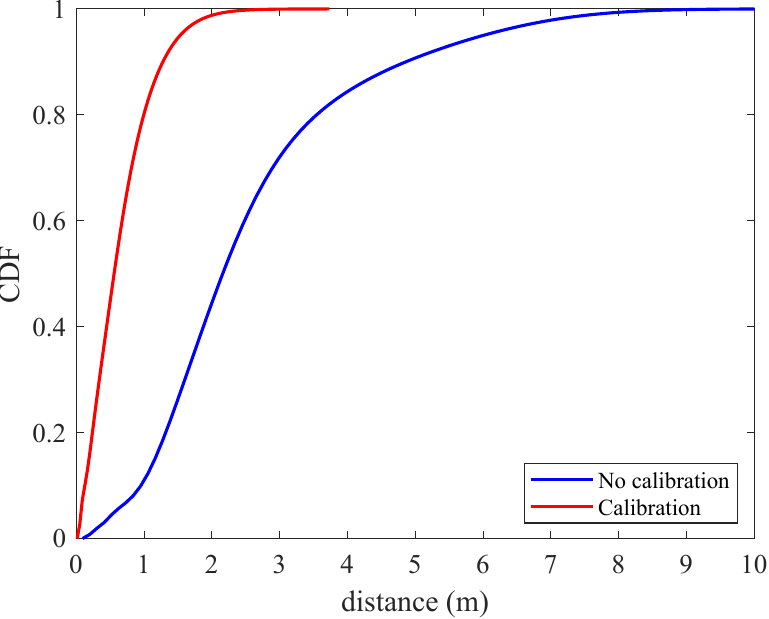}
\caption{Sensing location error result at a transmission power of 30 dBm.}
\label{fig_locerr}
\end{figure}

We conducted simulations under the fundamental parameters listed in Table \ref{tab:sim_param}. In addition, we set the channel generation interval to 0.625 ms, and the UE speeds are uniformly distributed between 3 and 4 m/s. Fig. \ref{fig_locerr} shows that applying the proposed calibration coefficients to correct for timing offset significantly improves localization accuracy compared to the uncalibrated system. Specifically, at a 90\% confidence level, the root mean square error (RMSE) is reduced by approximately 3 m.

\begin{figure}[t]
\centering
\includegraphics[width=0.65\linewidth]{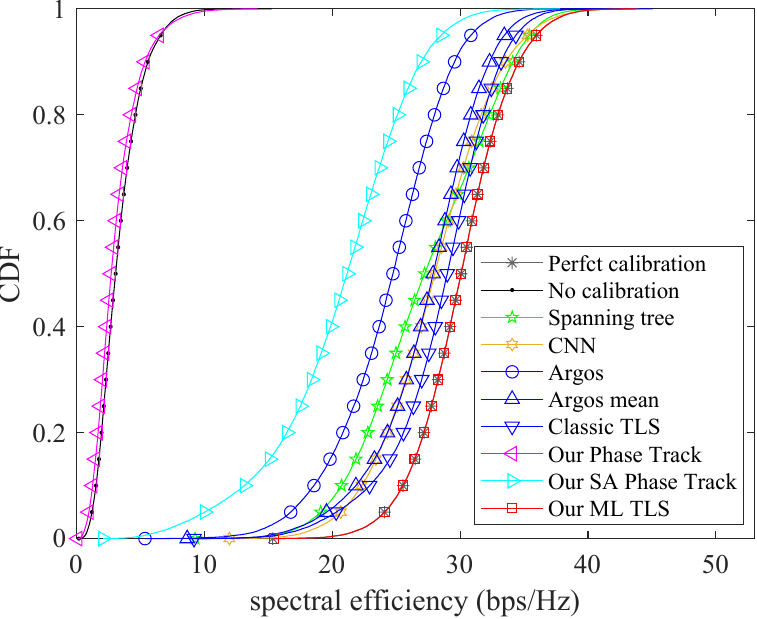}
\caption{Dynamic scene spectral efficiency result for different algorithms with the transmission power of 30 dBm.}
\label{fig_moveCDF}
\end{figure}

It can be observed from Fig. \ref{fig_moveCDF} that in dynamic scenes, the performance of the proposed two-step ML TLS algorithm applied per slot is close to the performance of perfect calibration. The performance is improved by 10 bps/Hz compared to the Argos algorithm applied per time slot. The proposed algorithm also outperforms both slot-by-slot Argos mean, TLS, spanning-tree and CNN calibration algoritmss. The performance of direct phase tracking in dynamic scenarios is similar to that of no tracking, with almost no cooperative transmission gain. By employing a sensing-assisted calibtation approach to reconstruct the OTA channel and then utilizing UL SRS channel phase tracking named SA phase tracking, the performance is nearly equivalent to that of the Argos algorithm. This demonstrates the effectiveness of sensing-assisted calibration, where the sensign results can assist in calibration, thereby reducing the calibration overhead. 

\begin{figure}[htbp]
\centering
\includegraphics[width=0.65\linewidth]{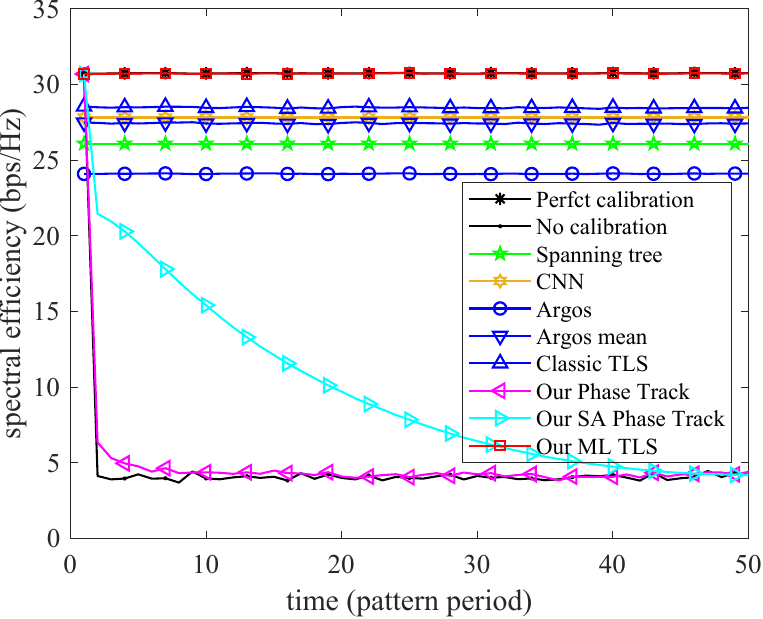}
\caption{The decay of spectrum efficiency over time with the transmission power of 30 dBm.}
\label{fig_changePeriod}
\end{figure}

Fig. \ref{fig_changePeriod} compares the performance of several calibration algorithms over time in a dynamic scenario, where phase tracking is conducted via the SRS channel in every TDD pattern. The TDD pattern in mmWave is typically 0.625 ms. The performance of direct phase tracking using the UL SRS is shown to be short-lived. It degrades rapidly, and the system loses all coherent transmission gain  by the fifth time slot. However, the SA phase track method proposed in this paper slows down the degradation of calibration performance. It loses coherent transmission gain totally at the $50$-th TDD pattern. In conclusion, the proposed SA phase track calibration demonstrates superior robustness. It maintains coherent gain for a much longer period.  Sensing-assisted calibration phase tracking can achieve coherent transmission performance gains with significantly lower overhead. In other words, the SA phase track calibration overhead can be reduced to $10\%$ compared to the previous method without sensing-assisted calibration.

\section{Experimental Measurement and Results}
\label{section5}
\subsection{Prototype System}
\begin{figure}[t]
\centering
\includegraphics[width=0.75\linewidth]{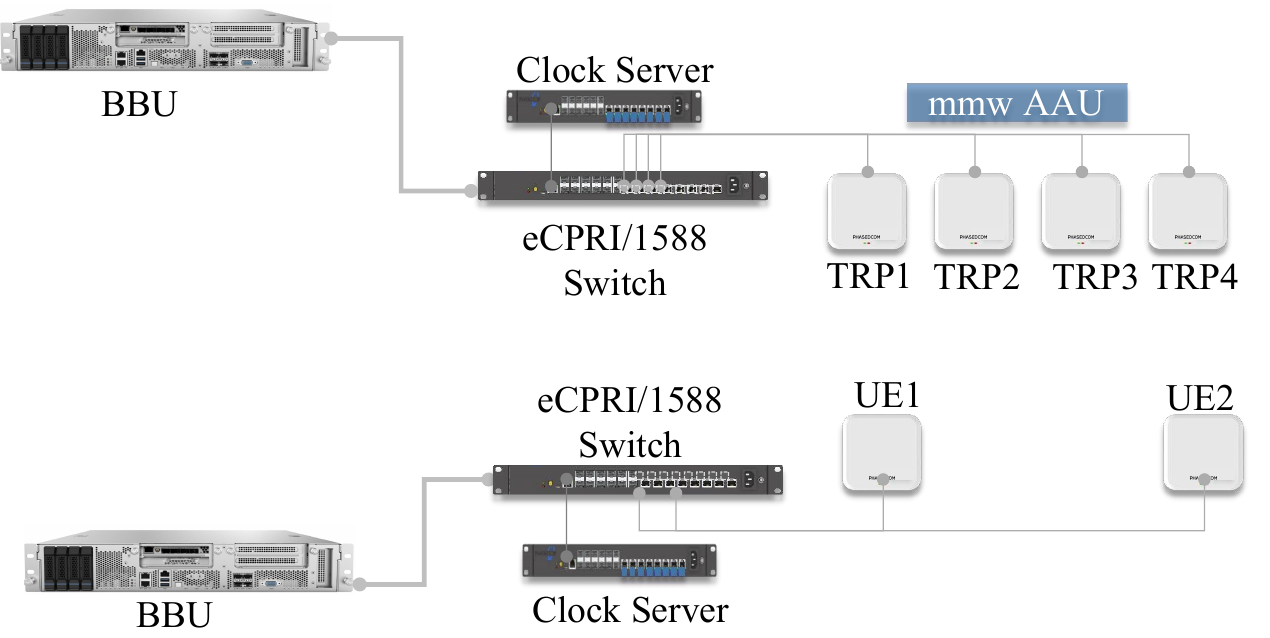}
\caption{Prototype system for coherent transmission and sensing.}
\label{fig_prototype}
\end{figure}

\begin{figure}[t]
\centering
\includegraphics[width=0.75\linewidth]{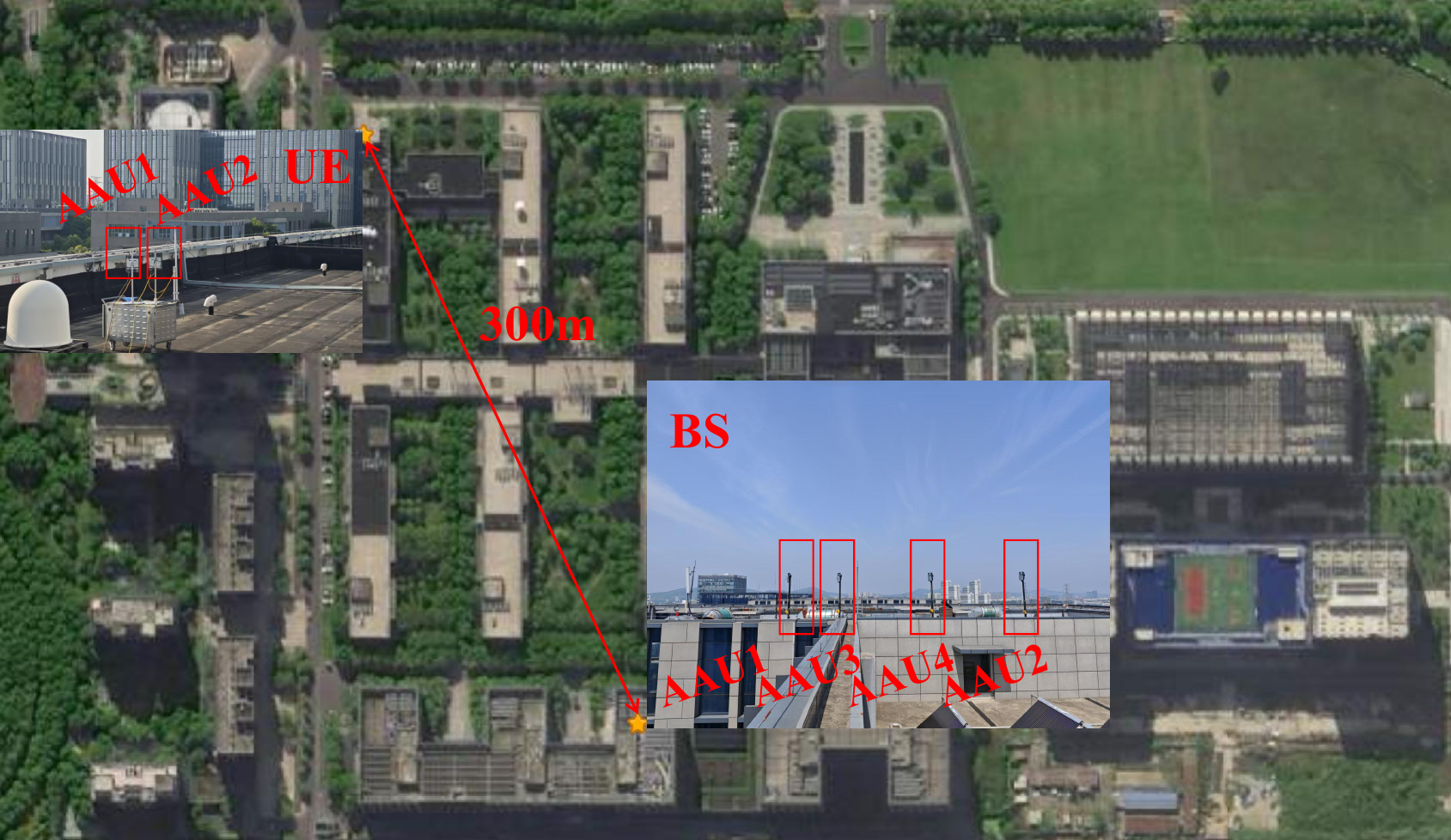}
\caption{Experimental scenario and system deployment diagram.}
\label{fig_exscene}
\end{figure}

We developed an experimental hardware platform for coherent transmission and sensing, which is shown in Fig. \ref{fig_prototype}. Commercial off-the-shelf (COTS) mmWave Active Antenna Units (AAUs) are used. The BS side is equipped with four AAUs and the UE side is equipped with two AAUs. Each AAU is configured with one carrier with 200 MHz bandwidth and features a dual-polarized antenna array with $4 \times 4$ antennas per array.  An evolved common public radio interface (eCPRI) is used between the AAU and the baseband unit (BBU), which complies with the open radio access network (ORAN). To handle the fronthaul throughput from the four AAUs, the BBU is equipped with two dual-port 25 Gbps Network Interface Cards (NICs). The clock server provides time synchronization and the reference clock for BBUs and AAUs through the switch by using IEEE 1588 precision time protocol (IEEE 1588 PTP) and synchronous Ethernet (SyncE) protocol. The system's frame structure adheres to the 3GPP standard, utilizing a subcarrier spacing of 120 kHz. As is shown in Fig. \ref{fig_exscene}, the experimental setup consisted of four TRPs with four AAUs and two prototype UEs with two AAUs. The coherent transmission and sensing experiment was conducted in an outdoor environment between two buildings, with the line-of-sight (LOS) path approximately 300 m. 

\subsection{Calibration Coefficients Characteristics}
\label{coefcharacter}
\begin{figure*}[htbp]
\centering
\subfloat[]{%
    \includegraphics[width=0.3\linewidth]{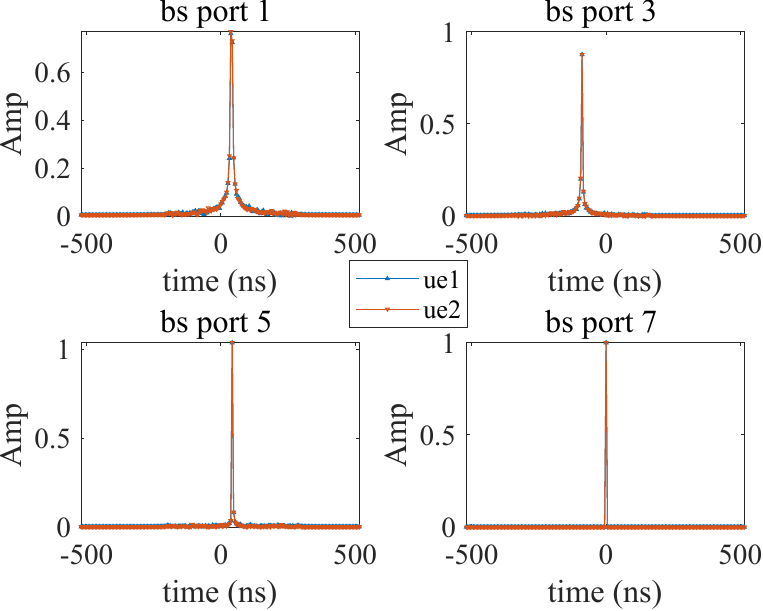}
    \label{fig_coef1ifft}}
\hfill
\subfloat[]{%
    \includegraphics[width=0.3\linewidth]{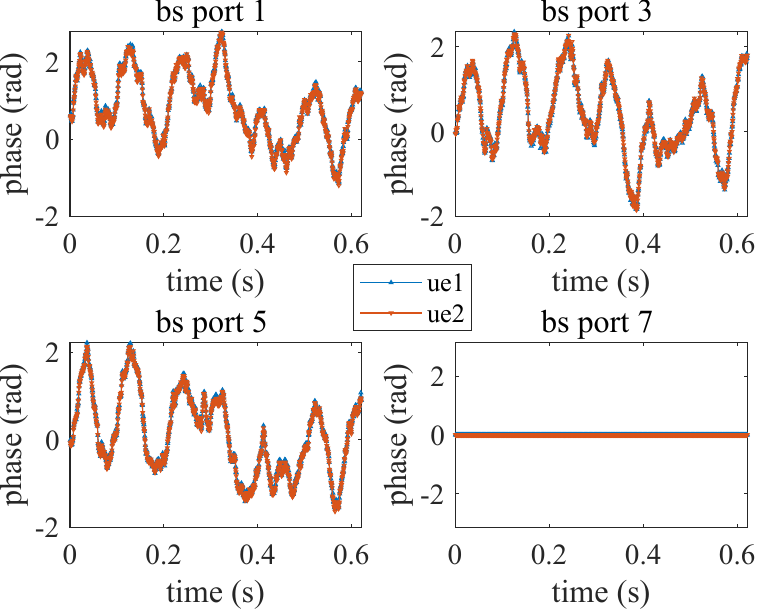}
    \label{fig_coef1phase}}
\hfill
\subfloat[]{%
    \includegraphics[width=0.3\linewidth]{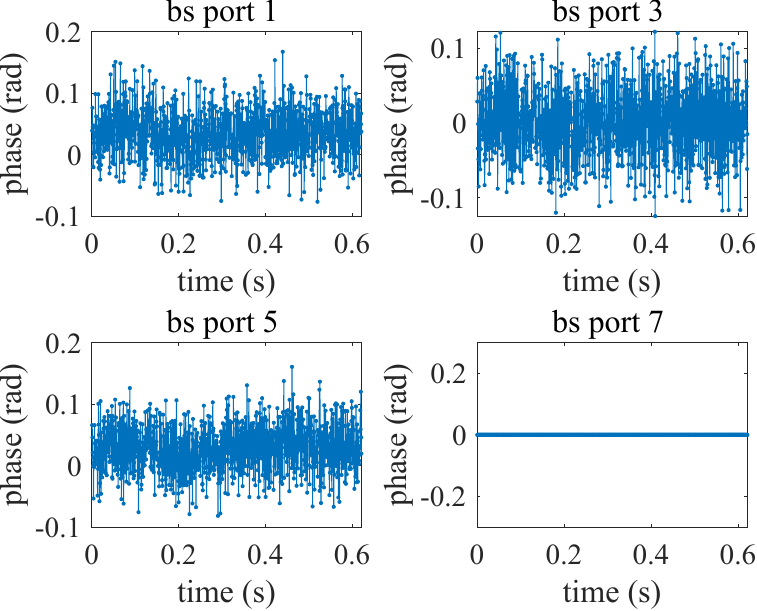}
    \label{fig_coef1phasediff}}
\vspace{0.1em}
\subfloat[]{%
    \includegraphics[width=0.3\linewidth]{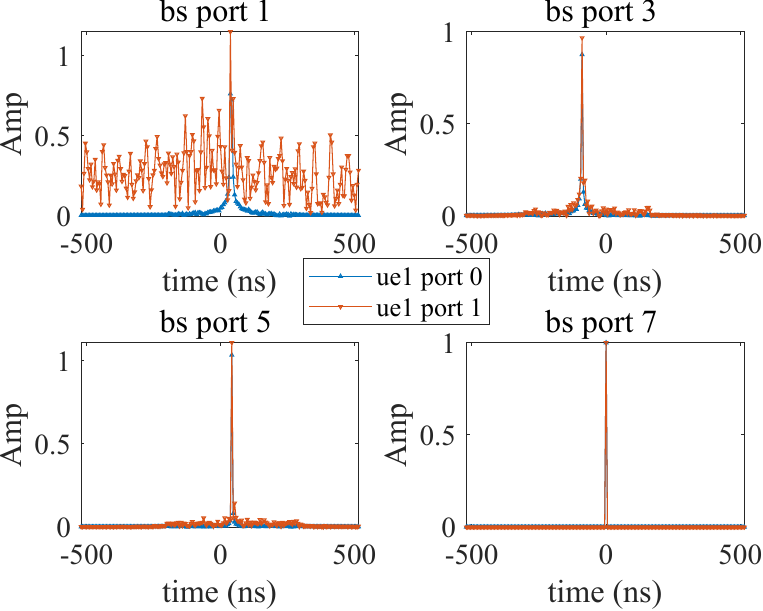}
    \label{fig_coef2ifft}}
\hfill
\subfloat[]{%
    \includegraphics[width=0.3\linewidth]{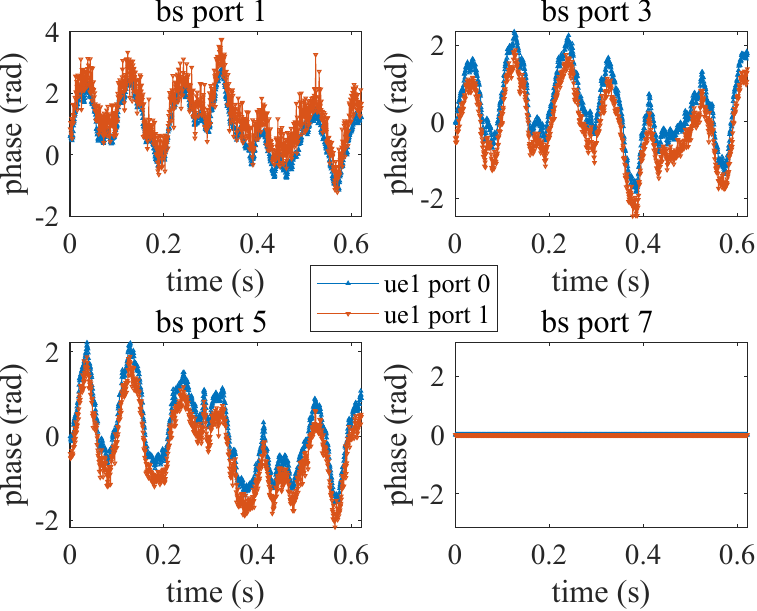}
    \label{fig_coef2phase}}
\hfill
\subfloat[]{%
    \includegraphics[width=0.3\linewidth]{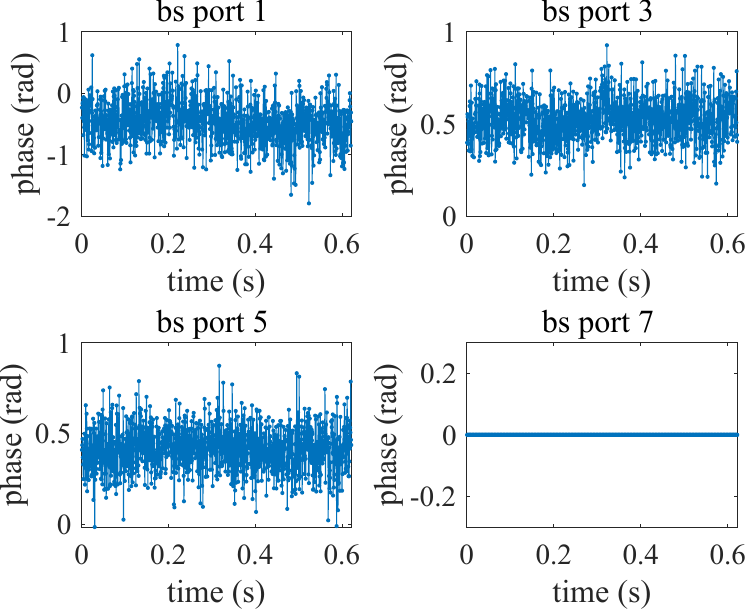}
    \label{fig_coef2phasediff}}
\caption{Characteristics of calibration coefficient normalized using the last port: (a) calibration delay of different UEs. (b) calibration phase trend of different UE. (c) calibration phase difference of different UE. (d) calibration delay of different polarization ports in the same UE. (e) calibration phase trend of different polarization ports in the same UE. (f) calibration phase difference of different polarization ports in the same UE.  }
\label{Fig1_coef}
\end{figure*}

First, we calculated the calibration coefficients per TDD pattern, which were normalized using the last port as a reference. The effect of different UEs on the polarization-matching port of the base station side is observed. As shown in Fig. \ref{fig_coef1ifft}, delay measurements of the calibration coefficients obtained from different UEs are consistent. As shown in Fig. \ref{fig_coef1phase}, the phase variation trend over time is consistent. As shown in Fig. \ref{fig_coef1phasediff}, the calibration coefficient phase measurements from different UEs are identical, meaning the phase difference is close to zero. Consequently, these findings validate that calibration coefficients measured using a single UE can be effectively applied to other UEs in a long-range deployment. This allows the system to achieve coherent transmission gain without requiring a separate calibration procedure for each UE.  
From Fig. \ref{fig_coef2ifft}, the measured delay can still be seen as the same. However, it is worth noting that the phase of the calibration coefficients measured from the two different polarization ports of the same UE is different shown in  Fig. \ref{fig_coef2phase} and Fig. \ref{fig_coef2phasediff}. The phase difference of different UE antenna ports deviates from zero. This discrepancy arises from signal polarization mismatch, which results in a low SNR for the cross-polarized channel. In such low-SNR conditions, noise significantly distorts the estimated phase of the coefficients. Therefore, obtaining accurate calibration coefficients depends on high SNR channels formed by polarization matching, while calibration coefficients measured from polarization-mismatched channels will be distorted.

\begin{figure}[htbp]
\centering
\subfloat[]{%
    \includegraphics[width=0.46\linewidth]{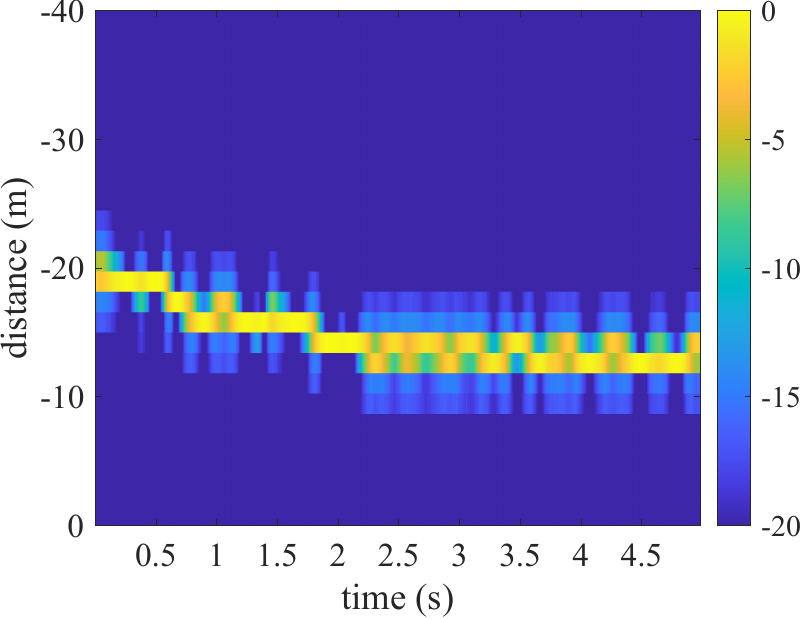}
    \label{fig_caliDis}}
\hfill
\subfloat[]{%
    \includegraphics[width=0.46\linewidth]{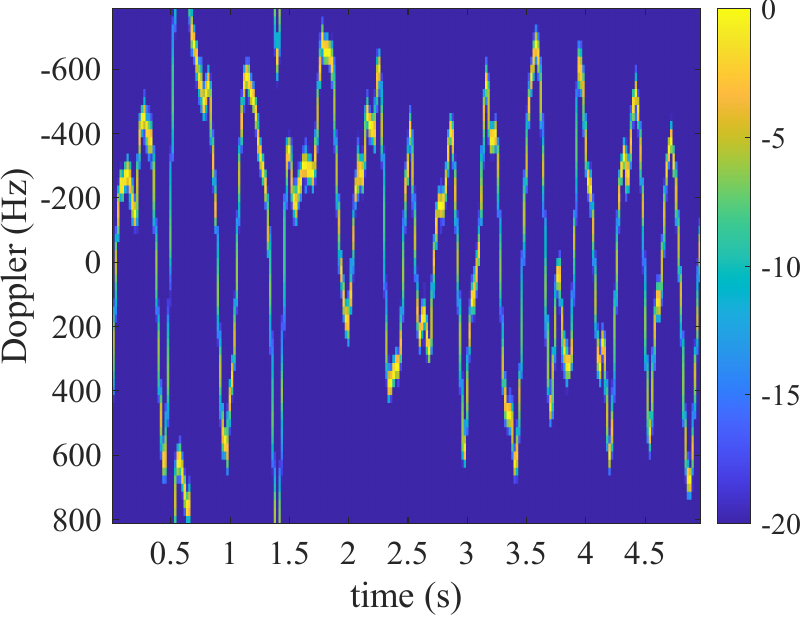}
    \label{fig_caliFre}}
\caption{Calibration coefficient timing and frequency offset variation over time. (a) timing offset variation which is change to distance; (b) frequency offset variation}
\label{Fig2_cali}
\end{figure}

The timing and frequency offset can be derived from the calibration coefficients obtained directly from the CARS feedback. According to our model, these derived offset should be double the values observed directly on the UL channel. Specifically, Fig. \ref{fig_caliDis} shows the timing offset changes about 6.3 m while  Fig. \ref{fig_ulDis} shows the timing offset changes about 3.2 m. It clearly demonstrates that the timing offset estimated from the calibration coefficients is approximately twice the timing offset measured on the UL channel. Likewise, Fig. \ref{fig_ulDis} shows the max frequency offset of about 550 Hz while Fig. \ref{fig_caliFre} shows the max frequency of about 1100 Hz  after unfolding. It validates that the estimated frequency offset is double the UL frequency offset.

\subsection{Communication Test Results}
\begin{figure}[htbp]
\centering
\includegraphics[width=0.65\linewidth]{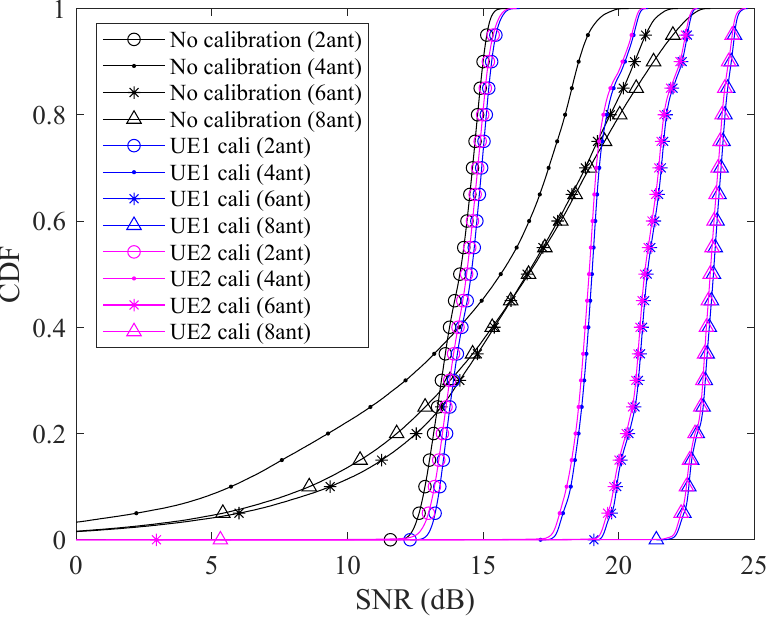}
\caption{The relationship between the single-stream DL SNR of the BS and the number of BS collaborative ports.}
\label{fig_bsS1SNR}
\end{figure}

The experiment began by generating two sets of calibration coefficients for the 8 ports of four TRPs at the BS side. The first set was derived from measurements using port 0 of UE1, and the second set was derived from port 0 of UE2. These coefficients were then used to evaluate the SNR of the single-stream DL transmission to UE1. The result is shown in Fig. \ref{fig_bsS1SNR}. First, in the uncalibrated baseline scenario, increasing the number of BS ports failed to provide a consistent SNR gain shown in black line. In contrast, when using the calibration coefficients of UE1, the SNR increases approximately 5 dB for each doubling of the number of ports, which is close to the theoretical 6 dB gain. This confirms proposed bidirectional calibration successfully enables DL coherent transmission. Besides, applying the calibration coefficients derived from UE2, it also yielded significant coherent gain which is slightly lower than that achieved with self-calibration. This crucial finding demonstrates that calibration coefficients are transferable between UEs, enabling coherent operation across the system with reduced calibration overhead.

\begin{figure}[htbp]
\centering
\includegraphics[width=0.65\linewidth]{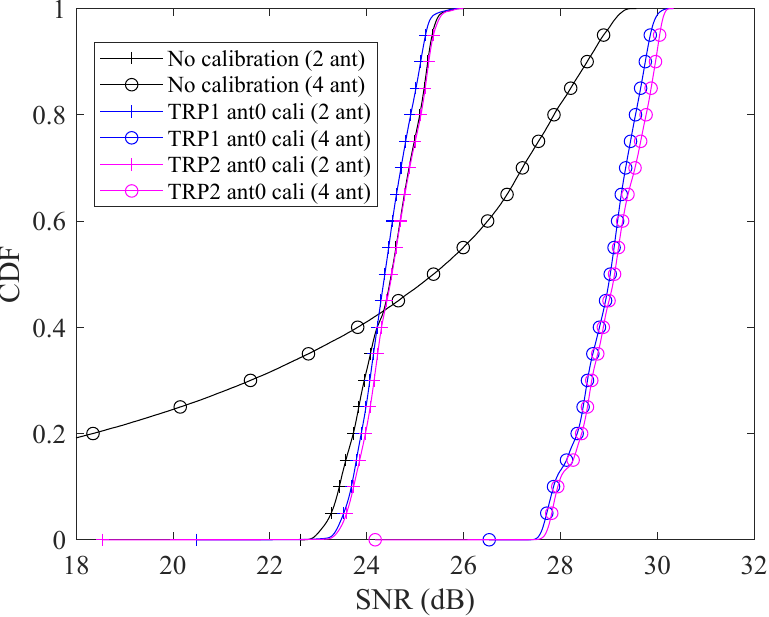}
\caption{The relationship between the single-stream UL SNR of collaborative UEs and the number of UE collaborative ports.}
\label{fig_ueS1SNR}
\end{figure}

Similarly, bidirectional calibration can also be performed to enhance single-stream UL SNR through UE collaboration, as demonstrated in Fig. \ref{fig_ueS1SNR}. To validate this, an experiment was conducted comparing a two-ports collaboration of UE1 with a four-ports collaboration of UE1 and UE2. Doubling the number of collaborative ports at the UE side yielded an SNR improvement of 4.6 dB shown in Fig. \ref{fig_ueS1SNR}. The results indicate that bidirectional calibration enables collaborative UE coherent transmission, which can extend the UE transmission distance.

\begin{figure}[t]
\centering
\includegraphics[width=0.65\linewidth]{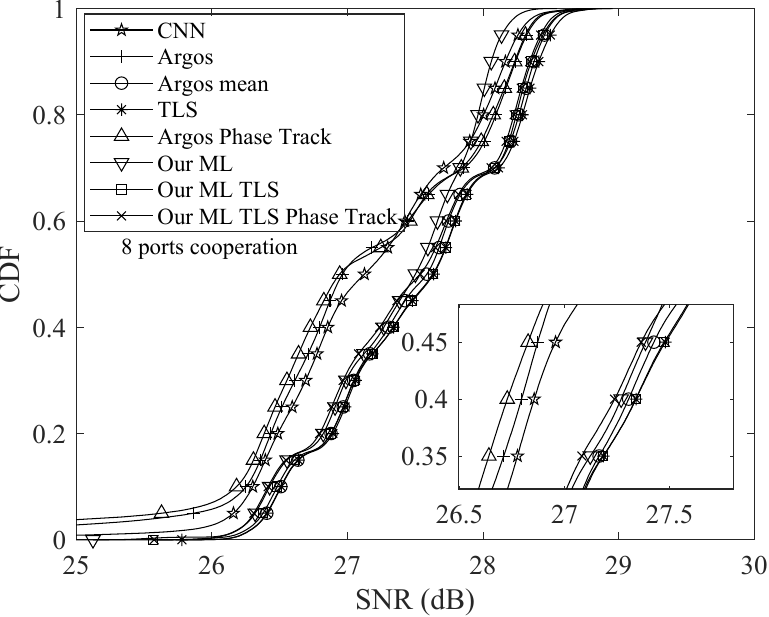}
\caption{Single-stream DL SNR of cooperative BSs using different calibration algorithms at 300 m.}
\label{fig_S1BSthms300m}
\end{figure}

\begin{figure}[t]
\centering
\includegraphics[width=0.65\linewidth]{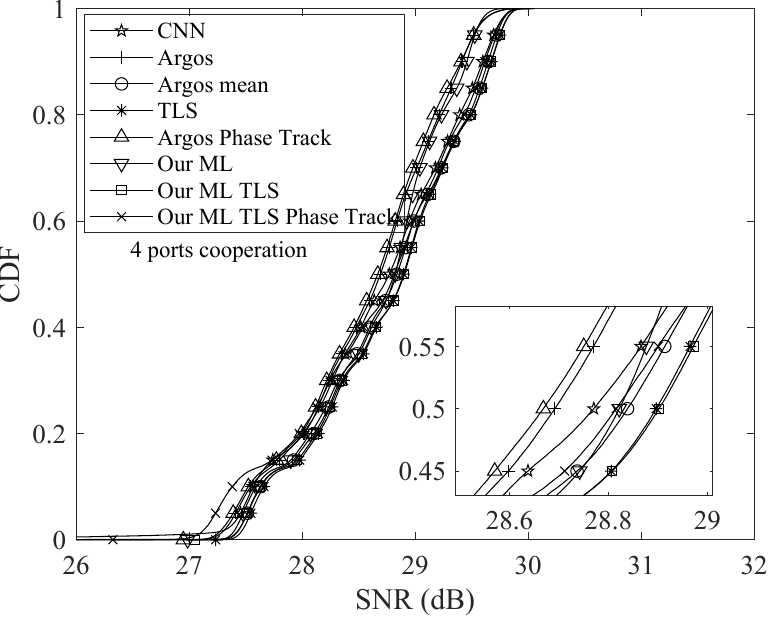}
\caption{Single-stream UL SNR of cooperative UEs using different calibration algorithms at 300 m.}
\label{fig_S1UEthms300m}
\end{figure}

Furthermore, Fig. \ref{fig_S1BSthms300m} presents a performance comparison of different calibration algorithms in DL transmission with four collaborative TRPs. Unlike the simulation, the collected channels are estimated. Therefore, we use the classic TLS algorithm applied per TDD pattern as the baseline for performance comparison. The results demonstrate that the proposed two-step ML TLS achieves a comparable level of performance. It outperforms classic Argos, Argos mean, TLS and CNN methods. The performance of the CNN-based calibration algorithm is superior to Argos, though slightly inferior to the phase tracking algorithm proposed in this paper. Furthermore, when combined with phase tracking, the proposed algorithm consistently outperforms the classic Argos algorithm per slot. Fig. \ref{fig_S1UEthms300m} presents a performance comparison of several calibration algorithms in UL transmission with two collaborative UEs. It shows that calibration algorithms combined with phase tracking algorithms exhibit some performance loss compared with using calibration algorithms per TDD pattern. However, the system can maintain UL coherent transmission using DL channel information to track and compensate for phase changes. All in all, Fig. \ref{fig_S1BSthms300m} and Fig. \ref{fig_S1UEthms300m} demonstrate that the proposed two-step ML TLS algorithm with phase tracking can provide a robust and low-overhead solution for practical deployments with achievable performance.

\begin{figure}[htbp]
\centering
\includegraphics[width=0.65\linewidth]{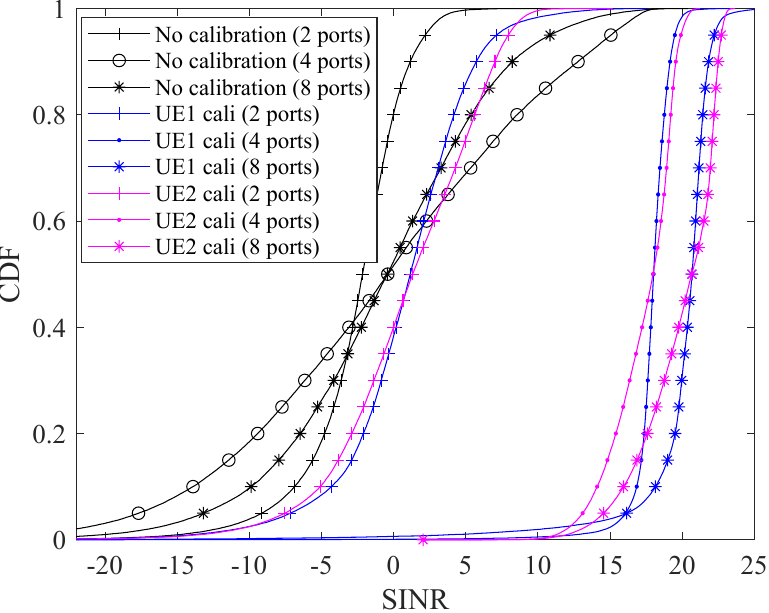}
\caption{Two-stream SINR of UE1 and UE2 with different number of BS collaborative ports.}
\label{fig_bsS2SINRdifferentue}
\end{figure}

We next investigated a two-stream DL transmission scenario, using signal-to-interference-plus-noise ratio (SINR) as the metric. Calibration coefficients for the 8-ports BS were generated using measurements from UE1 and UE2. The experiment shown in Fig. \ref{fig_bsS2SINRdifferentue} revealed that when only two BS ports were used to transmit the two streams, the resulting SNR was exceptionally low. This performance degradation is attributed to polarization matching. With only two ports, each UE effectively communicates with only one polarization-matched port, causing the multi-user channel matrix to become rank-deficient. A rank-1 channel cannot support two independent streams, leading to severe inter-user interference. In contrast, when the number of BS ports was increased to four, the channel rank became sufficient to support two streams, resolving the interference issue and generating a significant SNR gain. Furthermore, scaling the system from four to eight BS ports resulted in an additional coherent gain of approximately 3 dB. This confirms that once the channel rank is sufficient to support the number of streams, the system benefits from the expected coherent combining gain as more ports are added.

\begin{figure}[t]
\centering
\includegraphics[width=0.65\linewidth]{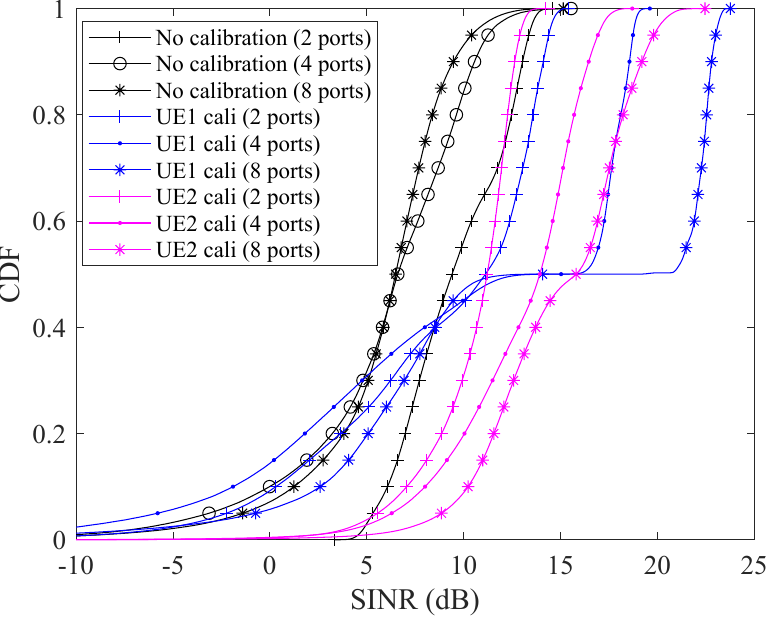}
\caption{Two-stream SINR of UE1  with different number of BS collaborative ports.}
\label{fig_bsS2SINRsingleue}
\end{figure}

We also examined the impact of using different ports from the same UE for calibration. Coefficients were generated separately using port 0 and port 1 of UE1. As discussed in section \ref{coefcharacter}, coefficients had identical time delays but distinct phase profiles. This phase difference led to highly asymmetric gains in a two-stream DL transmission, as illustrated in Fig. \ref{fig_bsS2SINRsingleue}. For instance, using coefficients of port 0 and doubling the number of BS ports yielded a 5 dB gain for one stream but only 1–2 dB for the other. This finding underscores that phase discrepancies between a UE's different ports can undermine multi-stream coherent gain, necessitating more advanced calibration methods.

\begin{figure}[htbp]
\centering
\includegraphics[width=0.65\linewidth]{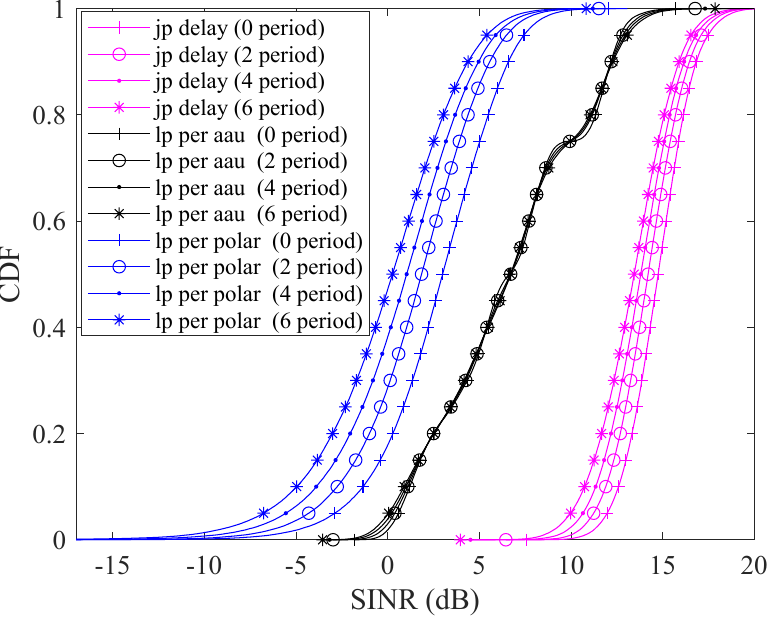}
\caption{Four-stream performance loss with calibration delay.}
\label{fig_thmsdelay}
\end{figure}

Taking into account computational and scheduling time in a practical system, we evaluated the impact of calibration delay on performance, with results shown in Fig. \ref{fig_thmsdelay}. The delay is quantified in units of a TDD period, where one period comprises five slots, totaling 0.625 ms. For a practical implementation delay of two TDD periods, the benchmark joint precoding scheme exhibits a minimal performance loss of approximately 0.5 dB. This loss increases with latency, reaching up to 2 dB for a delay of six TDD periods. The impact of this delay is more pronounced for local precoding schemes. The local precoding method per AAU demonstrates high resilience to delay, as the channels within a single AAU experience similar phase drift over time. In contrast, local precoding per polarization is far more sensitive, incurring a performance loss of approximately 1 dB at the two TDD-period delay.

\begin{figure}[htbp]
\centering
\includegraphics[width=0.65\linewidth]{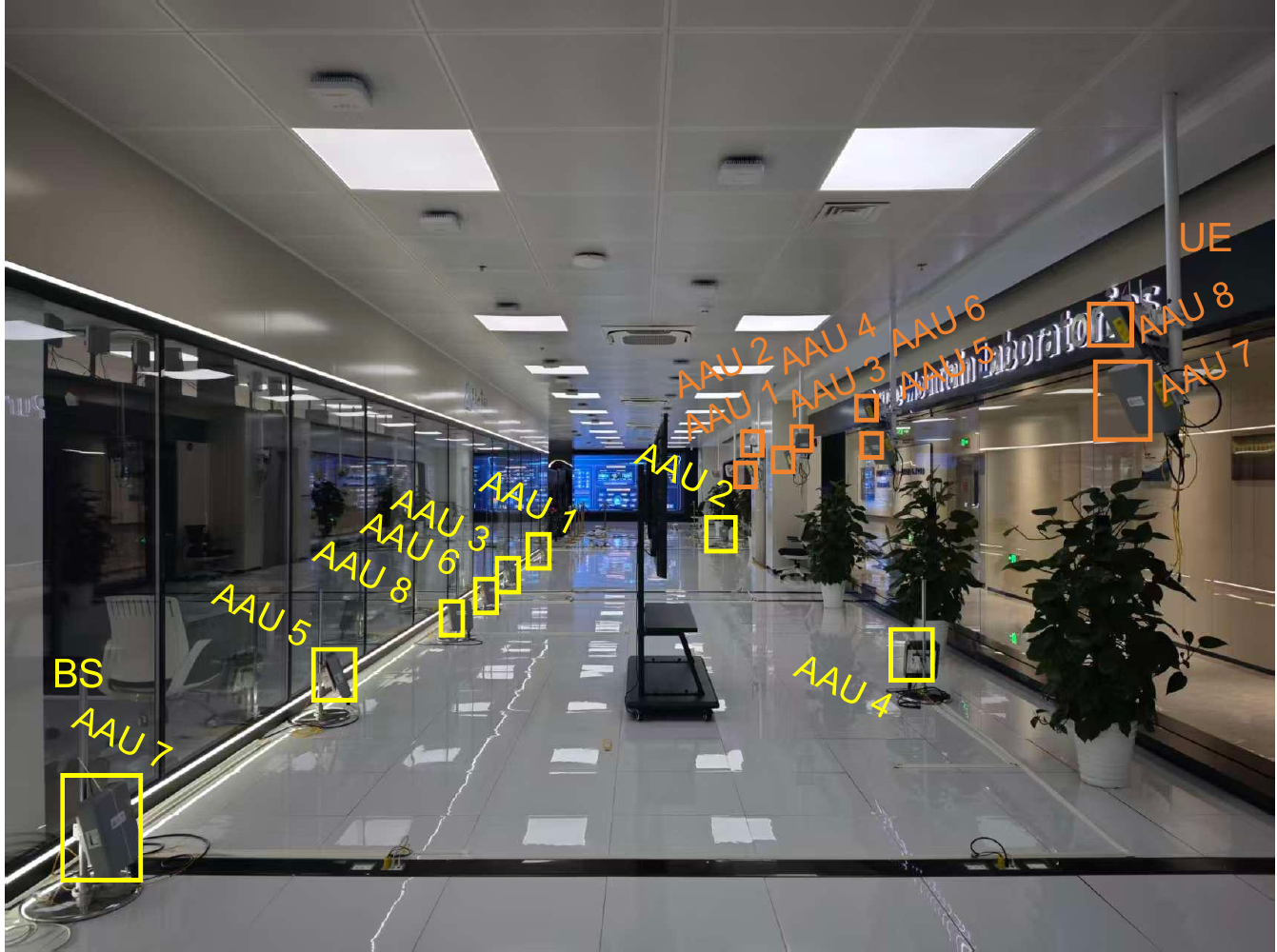}
\caption{Four-stream performance loss with calibration delay.}
\label{fig_bc1_scene2}
\end{figure}

To further evaluate the performance of the proposed algorithm in non-line-of-sight (NLOS) and rich multipath environments, we established a prototype system for indoor NLOS scenario testing. The experimental setup is illustrated in Fig. \ref{fig_bc1_scene2}, where 8 AAUs are deployed at the BS side and 8 AAUs at the UE side, with each AAU configuration consistent with the previous descriptions.

\begin{figure}[htbp]
\centering
\includegraphics[width=0.65\linewidth]{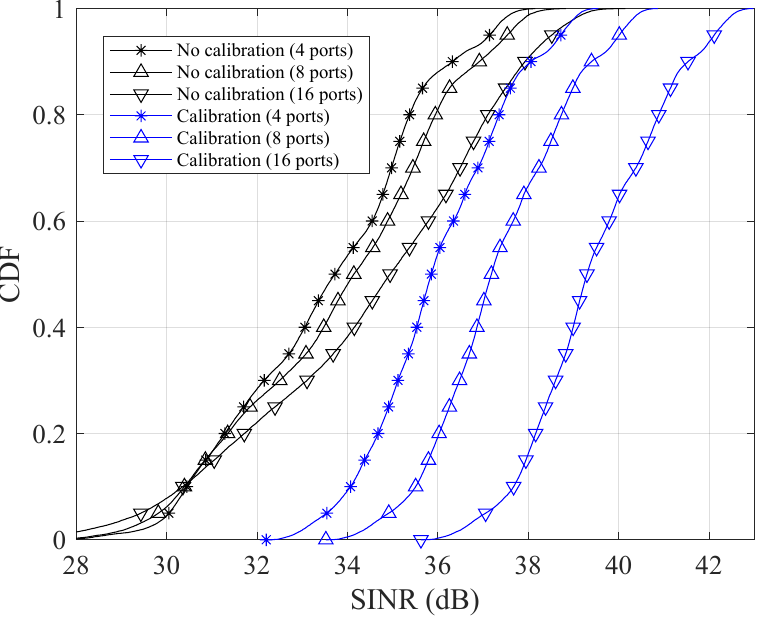}
\caption{Single-stream UE SNR versus the number of coordinated BS ports.}
\label{fig_bc2_SINR_Ants}
\end{figure}

First, we measured the average single-stream SNR for UEs at different locations in a single-stream configuration. The results, shown in Fig. \ref{fig_bc2_SINR_Ants}, indicate that without calibration, the downlink single-stream SNR at the BS side improves only marginally as the number of ports increases. In contrast, after applying the proposed calibration algorithm, the SNR increases significantly with the number of collaborative BS ports. This demonstrates that even in complex NLOS and rich multipath scenarios, calibration enables coherent cooperation at the BS side, thereby achieving SNR gains.

\begin{figure}[htbp]
\centering
\includegraphics[width=0.65\linewidth]{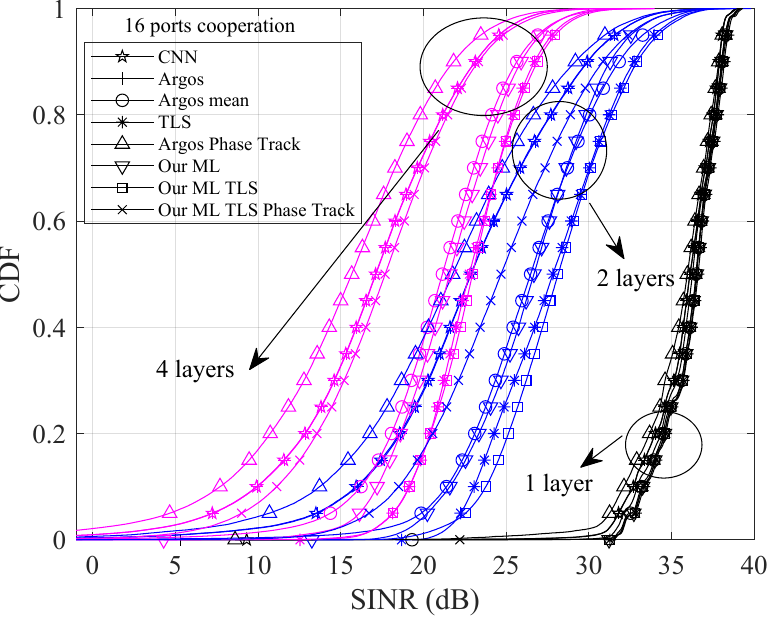}
\caption{SINR of cooperative BS ports using different calibration.}
\label{fig_bc3_SINR}
\end{figure}

We further compared the performance of different algorithms under varying numbers of streams, as depicted in Fig. \ref{fig_bc3_SINR}. First, across different stream configurations, the proposed ML TLS algorithm achieves slightly better performance than the conventional TLS method. The performance of the CNN-based calibration algorithm is comparable to that of Argos but lower than the phase tracking algorithm proposed in this scene. When using low-complexity phase tracking, the proposed algorithm outperforms the traditional Argos algorithm. Additionally, as the number of streams increases, the SINR per stream decreases, reflecting the trade-off between diversity and gain. A notable observation is that the performance gap between different algorithms becomes more pronounced with an increasing number of streams. The proposed calibration algorithm significantly improves the SINR in multi-stream scenarios, thereby enhancing overall system capacity.

\subsection{Sensing Test Results}
\begin{figure}[htbp]
\centering
\subfloat[]{%
    \includegraphics[width=0.46\linewidth]{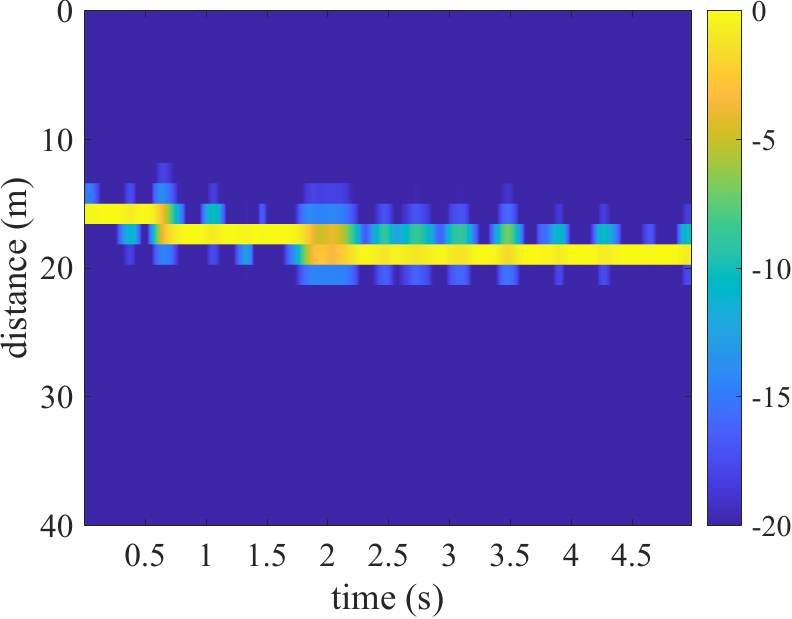}
    \label{fig_ulDis}}
\hfill
\subfloat[]{%
    \includegraphics[width=0.46\linewidth]{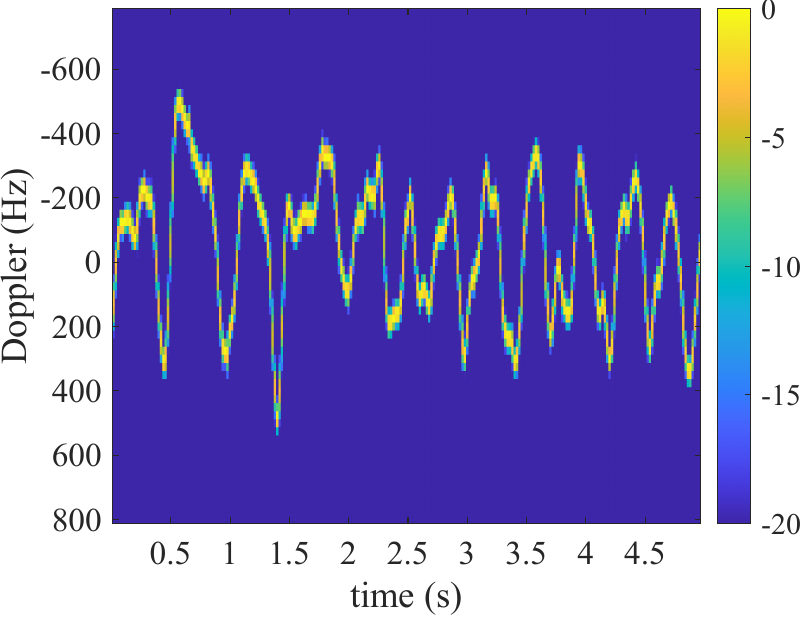}
    \label{fig_ulFre}}
\vspace{0.1em}
\subfloat[]{%
    \includegraphics[width=0.46\linewidth]{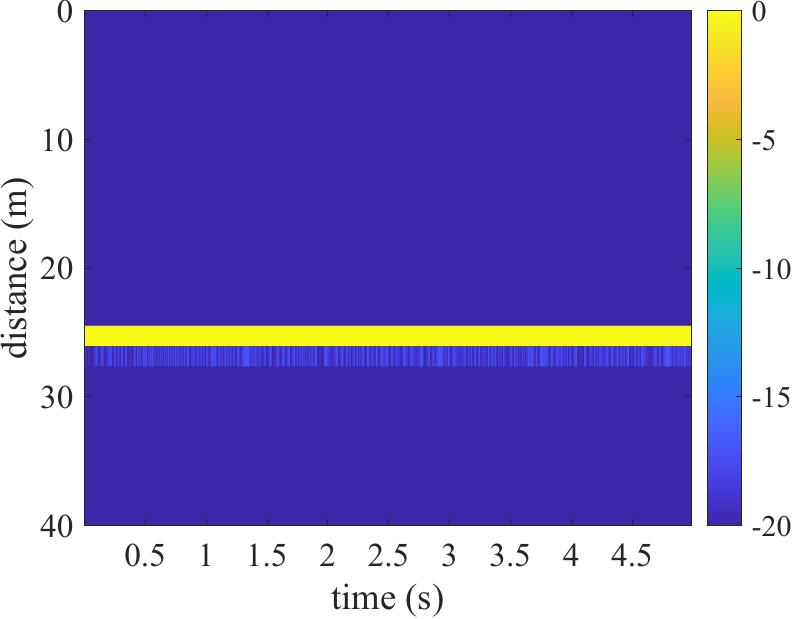}
    \label{fig_aftercaliDis}}
\hfill
\subfloat[]{%
    \includegraphics[width=0.46\linewidth]{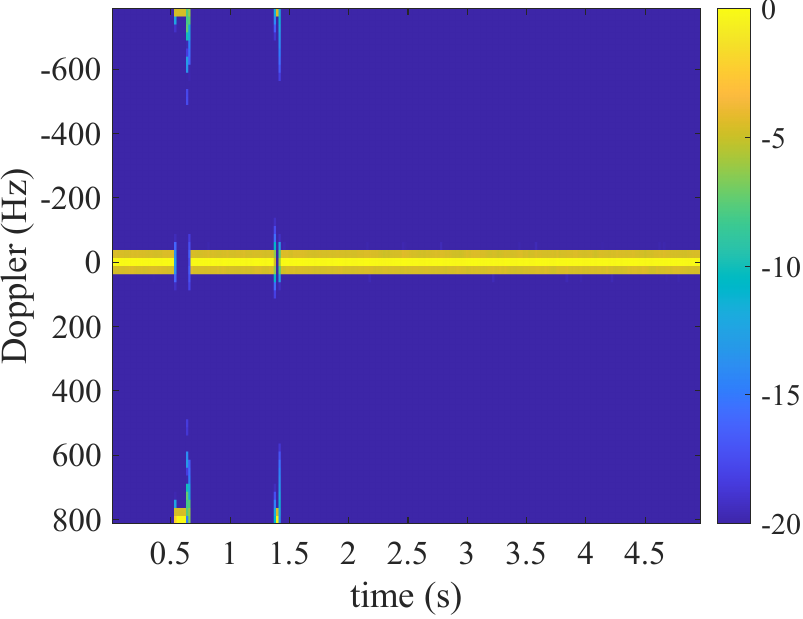}
    \label{fig_aftercaliFre}}
\caption{UL channel timing and frequency offset variation over time. (a) timing offset variation which is changed to distance before calibration; (b) frequency offset variation before calibration. (c) timing offset variation which is changed to distance  after calibration; (d) frequency offset variation after calibration.}
\label{Fig2_ul}
\end{figure}

This experiment took the UL sensing as an example, which is influenced by non-ideal factors between the BS and the UE. The results revealed two primary sources of instability: timing offset and frequency offset. As is shown in Fig.\ref{fig_ulDis}, the LOS path exhibited significant distance jitter of approximately 3.2 m over a representative 5-second measurement section, which corresponds to a timing offset of 10.6 ns. Simultaneously, a short-time fourier transform (STFT) analysis of the UL channel identified a dynamic frequency offset of up to 550 Hz shown in Fig. \ref{fig_ulFre}.

After calibrating the UL channel using reciprocity calibration, a clean OTA channel can be obtained, which eliminates the timing offset and frequency offset non-ideal factors. As is shown in Fig. \ref{fig_aftercaliDis}, the LOS path in the UL channel remains unchanged over the 5-second  observation window, indicating the elimination of timing offset non-ideal factors. As shown in Fig. \ref{fig_aftercaliFre}, the primary frequency of the frequency analysis of the UL channel is zero and remains constant, indicating the elimination of frequency offset in the LOS path. Another interesting phenomenon is that, at some certain moments, the calibrated frequency offset exhibits a jump. This occurs when non-ideal factors in the calibration coefficients exceed the sampling frequency of the calibration, causing aliasing and leading to calibration failure due to high-frequency offset. Therefore, to ensure robust and complete compensation, the sampling frequency of the calibration signal must be at least twice the maximum expected frequency deviation, in accordance with the Nyquist-Shannon sampling theorem.

\begin{figure}[t]
\centering
\subfloat[]{%
    \includegraphics[width=0.46\linewidth]{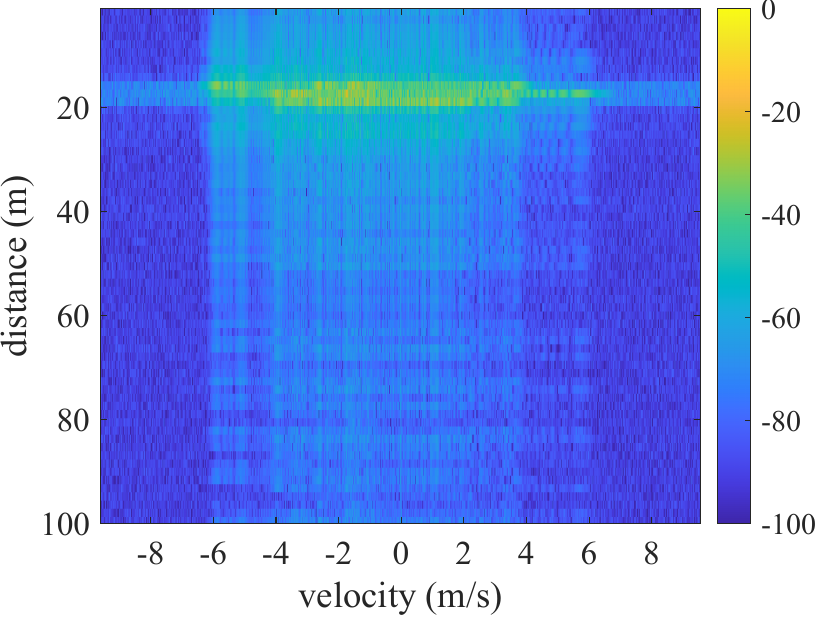}
    \label{fig_2dfftbefore}}
\hfill
\subfloat[]{%
    \includegraphics[width=0.46\linewidth]{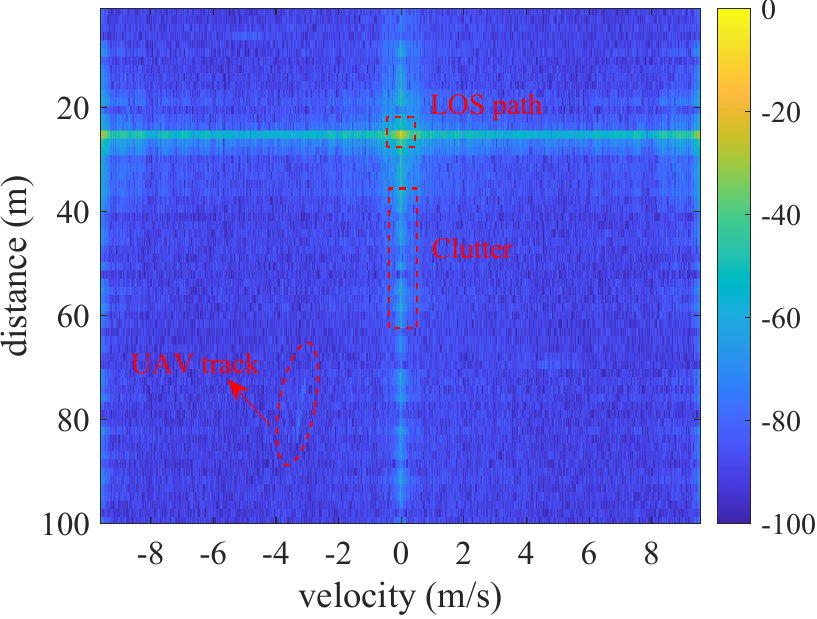}
    \label{fig_2dfftafter}}
\caption{(a) Range-Velocity Map before calibration. (b) Range-Velocity Map after calibration.}
\label{Fig2_2dfft}
\end{figure}

\begin{figure}[t]
\centering
\subfloat[]{%
    \includegraphics[width=0.46\linewidth]{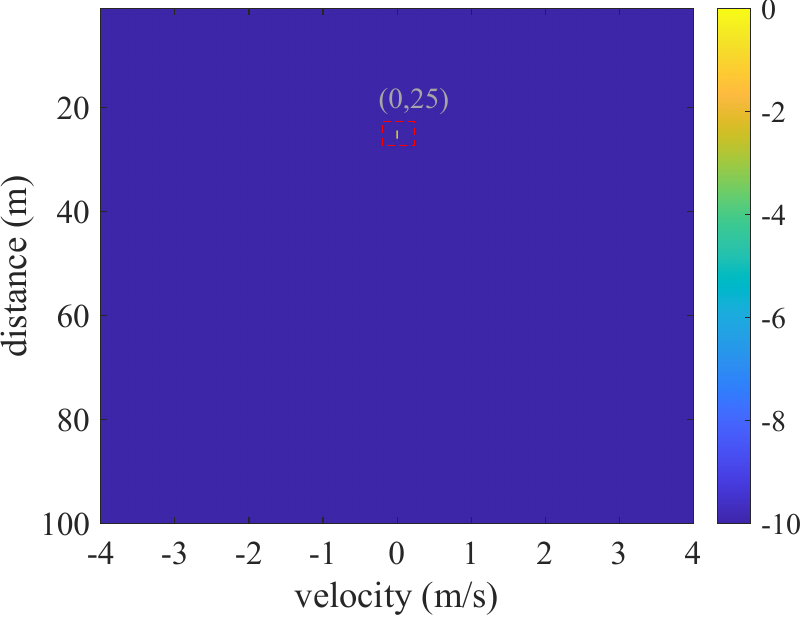}
    \label{fig_mtilos}}
\hfill
\subfloat[]{%
    \includegraphics[width=0.46\linewidth]{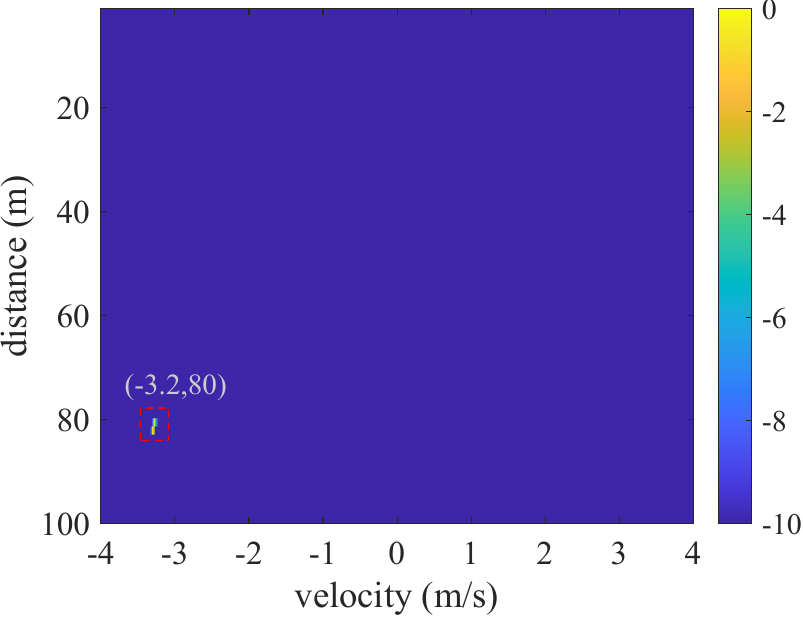}
    \label{fig_mtiuav}}
\caption{(a) Range-Velocity Detection before MTI. (b) Range-Velocity Detection after MTI.}
\label{Fig2_mti}
\end{figure}

The effect of calibration on channel sensing was evaluated using a conventional 2D-FFT algorithm. The uncalibrated channel, depicted in Fig. \ref{fig_2dfftbefore}, exhibits significant Doppler spread for the stationary LOS path. This apparent motion is an artifact caused by frequency offset variation between the transceivers, which introduces spurious Doppler shifts and violates the conditions required for coherent sensing. In contrast, after calibration, the sensing results in Fig. \ref{fig_2dfftafter} show a dramatic improvement. The frequency offset is eliminated, causing the LOS path energy to correctly collapse into a single, zero-Doppler peak.However, due to the large power difference between the LOS path and weak scatterers, the unmanned aerial vehicle (UAV) target remains weak and only LOS is detected in Fig. \ref{fig_mtilos}. Therefore, a moving target indication (MTI) algorithm\cite{Niu2025} is required to suppress these static components and reveal the dynamic target. We further applied the MTI algorithm to the calibrated channel. As shown in Fig. \ref{fig_mtiuav}, it successfully detected a moving target with a velocity of -3.2 m/s and a relative distance of 65 m from the LOS path. Given that the direct LOS path between the nodes was 300 m, the absolute distance of the UAV was calculated to be 365 m. The proposed calibration algorithm in this paper can effectively address non-ideal factors. However, to enable accurate phase tracking, the temporal density of sensing pilots must satisfy the Nyquist sampling criterion. Specifically, it must exceed twice the frequency offset between the transmitting and receiving nodes. Under this condition, the frequency offset can be accurately computed and compensated, thereby facilitating precise range estimation. The density of the calibrated time-frequency pilots determines the unambiguous range and velocity of the sensed target. Under the current low-density time-domain configuration, the maximum unambiguous velocity that can be achieved is 9.6 m/s, which represents the upper limit of measurable speed. Increasing the density of sensing pilots can further extend the perceivable velocity range. Moreover, since the calibration process relies on the reciprocity of the OTA channel, it is well-suited for multi-target scenarios and NLOS environments. All in all, it should be noted that achieving high-accuracy and unambiguous estimation of target range and velocity is constrained by two factors: the mitigation of non-ideal effects and the density of sensing pilots in the time-frequency domain. 

\section{Conclusion}
This paper presented a bidirectional calibration scheme forcoherent transmission and sensing, and developed a mmWave CF-mMIMO prototype platform based on COTS AAUs. We first designed a bidirectional calibration process and proposed a two-step ML TLS algorithm and phase track scheme to reduce calibration overhead. Furthermore, the scheme was extended to dynamic scenarios, where specific implementation methods for calibration-assisted sensing and sensing-assisted calibration were proposed to realize empowerment of both communication and sensing. Simulation results showed that sensing-assisted calibration can extend the coherence time of calibration, allowing for coherent transmission over a longer period, while calibration-assisted sensing improves the accuracy of sensing location. Experimental tests demonstrated that mmWave polarization-matched antennas can acquire the same calibration coefficients, whereas polarization-mismatched antennas introduce calibration coefficient deviations. For single-stream transmission, regardless of collaborative UE or collaborative TRP, each doubling of the number of cooperative antennas can achieve about 5 dB using the proposed calibration algorithm. For multi-stream transmission, the coherent transmission gain is subject to varying degrees of loss depending on the channel conditions. Sensing tests demonstrated that the calibrated channel enables coherent sensing over OFDM symbols, enabling high-precision velocity and distance sensing of long-range outdoor targets. Further research can be further conducted on low-overhead calibration in low SNR and high-speed mobility scenarios.
\label{section6}

\bibliographystyle{IEEEtran}
\bibliography{IEEEabrv,Bibliography}

\end{document}